\pdfoutput=1
\documentclass[
amsmath,
prd,nofootinbib,floatfix,11pt,
]{revtex4}

\usepackage{placeins}
\usepackage{afterpage}

\usepackage{axodraw2}
\usepackage{graphicx}
\usepackage{setspace}
\usepackage{bm}

\newcommand\beq{\begin{eqnarray}}
\newcommand\eeq{\end{eqnarray}}

\def\lsim{\mathrel{\rlap{\lower4pt\hbox{$\sim$}}
    \raise1pt\hbox{$<$}}}                
\def\gsim{\mathrel{\rlap{\lower4pt\hbox{$\sim$}}
    \raise1pt\hbox{$>$}}}            

\allowdisplaybreaks
\newcommand\lnbar{\overline{\ln}}
\newcommand\MSbar{$\overline{\rm{MS}}$ }
\newcommand\DRbar{$\overline{\rm{DR}}$ }
\newcommand\DRbarprime{$\overline{\rm{DR}}'$ }
\newcommand\Fbar{\overline F}

\begin{document}

\renewcommand{\theequation}{\arabic{section}.\arabic{equation}}
\renewcommand{\thefigure}{\arabic{section}.\arabic{figure}}
\renewcommand{\thetable}{\arabic{section}.\arabic{table}}

\title{\Large \baselineskip=20pt 
Three-loop effective potential for softly broken supersymmetry
}
\author{Stephen P.~Martin}
\affiliation{\mbox{\it Department of Physics, Northern Illinois University, DeKalb IL 60115}}

\begin{abstract}\normalsize \baselineskip=15.5pt 

The effective potential has been previously calculated through three-loop order, in Landau gauge, for a general renormalizable theory using dimensional regularization. However, dimensional regularization is not appropriate for softly broken supersymmetric gauge theories, because it explicitly violates supersymmetry. In this paper, I obtain the three-loop effective potential using a supersymmetric regulator based on dimensional reduction. Checks follow from the vanishing of the effective potential in examples with supersymmetric vacua, and from renormalization scale invariance in examples for which supersymmetry is broken, either spontaneously or explicitly by soft terms. As byproducts, I obtain the three-loop Landau gauge anomalous dimension for the scalar component of a chiral supermultiplet, and the beta function for the field-independent vacuum energy.

\end{abstract}

\maketitle

\tableofcontents

\baselineskip=14.7pt

\newpage

\section{Introduction\label{sec:intro}}
\setcounter{equation}{0}
\setcounter{figure}{0}
\setcounter{table}{0} 
\setcounter{footnote}{1}

The quantitative analysis of vacuum expectation values and spontaneous symmetry breaking
in quantum field theories can be formulated in terms of the Coleman-Weinberg effective potential \cite{Coleman:1973jx,Jackiw:1974cv,Sher:1988mj}. The perturbative loop expansion of the effective potential is evaluated as the the sum of all 1-particle irreducible vacuum
diagrams, where the vertices and propagators depend on the scalar background fields. For a general field theory, the effective potential is known at 2-loop order \cite{Ford:1992pn,Martin:2001vx} and 3-loop order \cite{Martin:2013gka,Martin:2017lqn}.
These results are based on Landau gauge fixing, which greatly simplifies the expressions; other gauge-fixing choices have kinetic mixing between scalar and vector degrees of freedom. Complete effective potential results for a general field theory at 2-loop order in a variety of other gauge-fixing prescriptions can be found in ref.~\cite{Martin:2018emo}, which illustrates the unfortunate complications encountered.
In the special case of the Standard Model, the 4-loop contributions at leading order in QCD are also known \cite{Martin:2015eia}.

The 3-loop effective potential results of ref.~\cite{Martin:2017lqn} were obtained using dimensional regularization (DREG) \cite{Bollini:1972ui,Bollini:1972bi,Ashmore:1972uj,Cicuta:1972jf,tHooft:1972tcz,tHooft:1973mfk} followed by renormalization with modified minimal subtraction, known as \MSbar \cite{Bardeen:1978yd,Braaten:1981dv}. Although \MSbar is the modern standard for loop calculations of all types in non-supersymmetric theories, it is not appropriate for supersymmetric theories with or without explicit soft breaking terms. This is because the DREG regularization procedure introduces explicit supersymmetry violation, due to the fact that in 
\beq
d = 4 - 2\epsilon
\eeq
dimensions there is a non-supersymmetric mismatch between the numbers of gauge boson and gaugino degrees of freedom. Although this mismatch only has multiplicity $2\epsilon$, it is multiplied by poles in $\epsilon$ from loop diagrams.
After renormalization, this leads to violations of the relationships among parameters that should be enforced by supersymmetry. 

The purpose of this paper is to remedy this problem by providing a counterpart to the results of ref.~\cite{Martin:2017lqn}, but using Siegel's supersymmetric regularization by dimensional reduction (DRED) \cite{Siegel:1979wq,Capper:1979ns,Jack:1997sr} followed by modified minimal subtraction.\footnote{Although there are technical problems \cite{Siegel:1980qs,Avdeev:1982xy,Stockinger:2005gx} associated with simultaneously avoiding either inconsistencies or ambiguities of DRED at higher loop orders while maintaining supersymmetry, these are not an issue for the 3-loop vacuum diagrams considered in this paper, as demonstrated by the explicit calculations reported below.} In DRED, loop momenta are still in $d$ dimensions, but each vector degree of freedom has 4 components, so as to avoid the non-supersymmetric mismatch between gauginos and gauge bosons. The extra $2\epsilon$ vector components are called $\epsilon$-scalars. When explicit soft supersymmetry breaking is present in supersymmetric gauge theories, there is an additional complication, because in general in the resulting \DRbar renormalization scheme the $\epsilon$-scalars obtain non-zero squared mass contributions in excess of the corresponding vector squared masses, due to renormalization. These $\epsilon$-scalar squared mass contributions are unphysical, in the sense that they have no observable counterparts. Accordingly, in ref.~\cite{Jack:1994rk} it was shown that these unphysical quantities can be simultaneously eliminated from the renormalization group equations and from the relations between on-shell physical quantities and the Lagrangian parameters, by a parameter redefinition of the type given in ref.~\cite{Jack:1994kd}. The resulting supersymmetric renormalization scheme based on regularization by dimensional reduction is known\footnote{Many sources elide the distinction between the \DRbarprime and \DRbar schemes. It is hard to fault this practice, as the \DRbar scheme as defined in ref.~\cite{Jack:1994rk} and in the present paper (including arbitrary independent unphysical $\epsilon$-scalar squared masses) is not of much practical use.} as the \DRbarprime scheme, with the property that the $\epsilon$-scalar squared masses appearing in propagators are exactly the same as those of the corresponding vector bosons. The 2-loop results for the effective potential in a general softly broken supersymmetric gauge theory in \DRbarprime were obtained in ref.~\cite{Martin:2001vx}, and in the present paper this will be extended to 3-loop order.

The notations, conventions, and general strategies of this paper will follow closely those of refs.~\cite{Martin:2017lqn} and \cite{Martin:2016bgz}. Therefore, to avoid needless (and lengthy) repetition,
the reader is advised to consult those papers for the relevant definitions. In particular, the 3-loop effective potential is given in terms of renormalized $\epsilon$-finite basis integrals: 
$A(x)$ at 1-loop, $I(x,y,z)$ at 2-loops, and $F(w,x,y,z)$, $\overline F(0,x,y,z)$, $G(v,w,x,y,z)$, and $H(u,v,w,x,y,z)$ at 3-loops, along with convenient combinations $\overline A(x,y)$, $\overline I(w,x,y,z)$, and $K(u,v,w,x,y,z)$. Here $u,v,w,x,y,z$ denote propagator squared mass arguments, and the dependence on the common renormalization scale $Q$ is suppressed in the lists of arguments, as it is typically the same everywhere within a given calculation. 
These basis functions were defined explicitly in section II of \cite{Martin:2017lqn} and section II of \cite{Martin:2016bgz}, and the computer software library {\tt 3VIL} provided with the latter reference provides for their fast and accurate numerical evaluation. Note that their definitions do not depend on whether one is using the \MSbar or \DRbarprime scheme.
They satisfy symmetry relations that reflect all of the invariances of the corresponding underlying Feynman diagrams under interchanges of squared mass arguments. They also satisfy special case relations, which are identities that occur when the squared mass arguments are non-generic, meaning that some of them are equal to each other, and/or vanish. Examples of these special case relations appeared in eqs.~(5.82)-(5.86) of ref.~\cite{Martin:2016bgz} and (2.40)-(2.43) of ref.~\cite{Martin:2017lqn}. There are many other identities reflecting the analytic special cases that occur when there is only one distinct non-zero squared mass, found in 
refs.~\cite{Broadhurst:1991fi,Avdeev:1994db,Fleischer:1994dc,Avdeev:1995eu,Broadhurst:1998rz,Fleischer:1999mp,Schroder:2005va}, and listed in the notation of the present paper in section V of ref.~\cite{Martin:2016bgz}. For convenience, both the symmetry relations and the known special case relations are collected in an ancillary file {\tt identities.anc} distributed with the present paper. 

Since the structure of the 3-loop effective potential has been elucidated already in ref.~\cite{Martin:2017lqn}, at considerable length, the present paper will assume this as given, and concentrate on the distinctions that are special to supersymmetric theories and DRED. Furthermore, the explicit results at 3-loop order are extremely complicated, and therefore mostly useless to the human eye. Therefore, they will be almost entirely relegated to ancillary electronic files, which are suitable for use with symbolic manipulation software and numerical evaluation with {\tt 3VIL}.

\section{Effective potential in dimensional reduction\label{sec:effpotDRED}}
\setcounter{equation}{0}
\setcounter{figure}{0}
\setcounter{table}{0} 
\setcounter{footnote}{1}

Consider a general renormalizable theory, which we will later assume to be a softly broken supersymmetric gauge theory. Suppose that the fields with diagonal tree-level squared masses consist of some real scalars $R_j$ with squared masses $m^2_j$, two-component fermions $\psi_I$ with squared masses $M_I^2$, and 
real vector fields $A_a^\mu$ with squared masses $m_a^2$. In the case of the fermions, the masses need not be diagonal,
but may include charged Dirac fermion fields consisting of pairs $\psi_I$ and $\psi_{I'}$ with off-diagonal masses $M^{II'}$, where $M_I^2 = M_{I'}^2 = |M^{II'}|^2$. For Majorana fermions, one identifies $I$ and $I'$. There are also field-dependent interactions 
\beq
{\cal L}_{\rm int} &=&
-\frac{1}{6} \lambda^{jkl} R_j R_k R_l
-\frac{1}{24} \lambda^{jklm} R_j R_k R_l R_m
-\frac{1}{2} \left ( Y^{jIJ} R_j \psi_I \psi_J + {\rm c.c.} \right )
\nonumber \\ &&
+ g^{aJ}_I A^{\mu a} \psi^{\dagger I} \overline{\sigma}_\mu \psi_J
- g^{ajk} A^{\mu a}R_j \partial_\mu R_k
- \frac{1}{4} g^{abjk} A_\mu^a A^{\mu b} R_j R_k
- \frac{1}{2} g^{abj} A_\mu^a A^{\mu b} R_j
\nonumber \\ &&
- g^{abc} A^{\mu a} A^{\nu b} \partial_\mu A_\nu^c
- \frac{1}{4} g^{abe} g^{cde} A^{\mu a} A^{\nu b} A^c_\mu A^d_\nu
- g^{abc} A^{\mu a} \omega^b \partial_\mu \overline{\omega}^c
,
\label{eq:Linteractions}
\eeq
where $\omega^a$ and $\overline{\omega}^c$ are ghost and anti-ghost fields.
The independent couplings are scalar cubic $\lambda^{jkl}$, scalar quartic $\lambda^{jklm}$, Yukawa $Y^{jIJ}$,
vector-fermion-fermion $g^{aJ}_I$, vector-scalar-scalar $g^{ajk}$, vector-vector-scalar $g^{abj}$, and vector-vector-vector $g^{abc}$. By convention, $Y_{jIJ} \equiv (Y^{jIJ})^*$ and $M_{II'} = (M^{II'})^*$.  Note that the vector-vector-scalar-scalar and vector-vector-vector-vector interaction couplings are not independent of the cubic couplings, as they are given by
\beq
g^{abjk} &=& g^{ajl} g^{bkl} + g^{akl} g^{bjl}
,
\label{eq:gabjk}
\\
g^{abcd} &=& g^{abe} g^{cde}
,
\eeq
respectively. Each of these masses and couplings
may depend on one or more background scalar fields $\varphi$, which correspond to the possible vacuum expectation values.

The loop expansion of the \DRbarprime effective potential can be written as
\beq
V_{\rm eff} &=& V^{(0)} + \frac{1}{16\pi^2} V^{(1)} + \frac{1}{(16\pi^2)^2} V^{(2)} + \frac{1}{(16\pi^2)^3} V^{(3)} + \ldots
.
\eeq
The contribution $V^{(0)}$ is the tree-level background-field-dependent potential, and each $V^{(\ell)}$ is obtained
by summing the contributions of $\ell$-loop 1-particle-irreducible diagrams.
At 1-loop order, 
the effective potential in the \DRbarprime scheme is given by the supertrace form
\beq
V^{(1)} &=& \sum_j f(j) - 2 \sum_I f(I) + 3 \sum_a f(a),
\label{eq:oneloopV}
\eeq
where $j$, $I$, and $a$ appearing as arguments of loop integral functions are short-hand notations for the corresponding \DRbarprime squared masses, and the 1-loop integral function is
\beq
f(x) &=& \frac{1}{4} x^2 (\lnbar(x) - 3/2),
\eeq
which depends on the renormalization scale $Q$ through the definition
\beq
\lnbar(x) = \ln(x/Q^2).
\eeq
As explained in ref.~\cite{Martin:2001vx}, 
eq.~(\ref{eq:oneloopV}) differs from the \MSbar result, which instead has a 1-loop function $f_V(x) = f(x) + x^2/6$ for the vectors. The difference arises from the $\epsilon$-scalar contribution to $f(x)$.

The two-loop contribution, in either \MSbar or $\overline{\rm DR}'$, can be written in the form
\beq
V^{(2)} &=&
\frac{1}{12} (\lambda^{jkl})^2 f_{SSS}(j,k,l)
  +\frac{1}{8} \lambda^{jjkk} f_{SS}(j,k)
\nonumber \\ &&
  +\frac{1}{2} Y^{jIJ} Y_{jIJ} f_{FFS}(I,J,j)
  +\frac{1}{4} \left (Y^{jIJ} Y^{jI'J'} M_{II'} M_{JJ'} + {\rm c.c.} \right )
  f_{\Fbar\Fbar S}(I,J,j)
\nonumber \\ &&
  +\frac{1}{4} (g^{ajk} )^2  f_{VSS}(a,j,k)
  +\frac{1}{4} (g^{abj} )^2  f_{VVS}(a,b,j)
\nonumber \\ &&
  +\frac{1}{2} g^{aJ}_I g^{aI}_J f_{FFV}(I,J,a)
  +\frac{1}{2} g^{aJ}_I g^{aJ'}_{I'} M^{II'} M_{JJ'} f_{\Fbar\Fbar V}(I,J,a)
\nonumber \\ &&
  + \frac{1}{12} (g^{abc})^2 f_{\mbox{gauge}}(a,b,c)
,
\label{eq:V2loopgeneral}
\eeq
in terms of two-loop integral functions
$f_{SSS}$, $f_{SS}$, $f_{FFS}$, $f_{\Fbar\Fbar S}$,
$f_{VSS}$, $f_{VVS}$, $f_{FFV}$, $f_{\Fbar\Fbar V}$, and $f_{\rm gauge}$. The functions $f_{SSS}$, $f_{SS}$, $f_{FFS}$,
and $f_{\Fbar\Fbar S}$ do not involve vectors or $\epsilon$-scalars, and so
are trivially the same in the \MSbar and \DRbarprime schemes. In contrast,
the functions $f_{VSS}$, $f_{VVS}$, $f_{FFV}$, $f_{\Fbar\Fbar V}$, and $f_{\rm gauge}$ are different in the two schemes.
The \DRbarprime functions are constructed so as to include the contributions of the $\epsilon$-scalars corresponding to each vector field, with each $\epsilon$-scalar mass equal to the corresponding field-dependent vector boson mass. They were obtained in\footnote{Here we have adopted a slightly more efficient notation than in that paper, since 
$f_{VSS}(x,y,z) \equiv f_{SSV}(y,z,x) + F_{VS}(x,y) + F_{VS}(x,z)$,
where the functions on the right side were the ones defined in ref.~\cite{Martin:2001vx}, and the function $f_{VSS}$
is the one used here. This takes advantage of eq.~(\ref{eq:gabjk}).} 
ref.~\cite{Martin:2001vx}.
The \DRbarprime results for the one-loop function $f$ and the 9 two-loop functions are provided in the ancillary file
{\tt functionsDRED.anc} provided with this paper. 

\begin{figure}[t]
\begin{center}
\includegraphics[width=0.99\linewidth,angle=0]{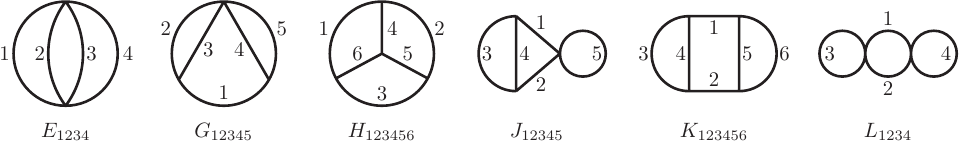}
\end{center}
\vspace{-0.25cm}
\begin{minipage}[]{0.96\linewidth}
\caption{\label{fig:diagramtopologies}
Feynman diagram topologies that contribute to the effective potential at 3-loop order. The numbers
indicate the canonical ordering of subscripts denoting propagator
types ($S$, $F$, $\Fbar$, $V$), and the corresponding squared mass arguments.}
\end{minipage}
\end{figure}

As explained in ref.~\cite{Martin:2017lqn}, the 3-loop contribution to the effective potential for a general renormalizable
theory can be expressed in terms of 89 loop integral functions; see eqs.~(3.2)-(3.32) of that paper for the rather lengthy expression for $V^{(3)}$ in terms of the functions and the renormalized couplings. The 89 functions can be divided into three categories. First, there are 24 functions that do not involve vector fields or $\epsilon$-scalars at all, and so are trivially the
same in the \DRbarprime and \MSbar schemes: 
\beq
&& \!\!\!\!\!\!\!
H_{SSSSSS},\>
K_{SSSSSS},\>
J_{SSSSS},\>
G_{SSSSS},\>
L_{SSSS},\>
E_{SSSS},\>
\nonumber \\ && \!\!\!\!\!\!\!
H_{FF\Fbar SSS},\>
H_{\Fbar\Fbar\Fbar SSS},\>
H_{FFSSFF},\>
H_{FFSS\Fbar\Fbar},\>
H_{F\Fbar SSF\Fbar},\>
H_{\Fbar\Fbar SS\Fbar\Fbar},\>
\nonumber \\ && \!\!\!\!\!\!\!
K_{SSSSFF},\>
K_{SSSS\Fbar\Fbar},\>
K_{FFFSSF},\>                                                                  
K_{FF\Fbar SS\Fbar},\>
K_{\Fbar\Fbar FSSF},\>
K_{\Fbar F\Fbar SSF},\>                                                        
\nonumber \\ && \!\!\!\!\!\!\!
K_{\Fbar\Fbar\Fbar SS\Fbar},\>                                                 
K_{SSFFFF},\>
K_{SSFF\Fbar\Fbar},\>
K_{SS\Fbar\Fbar\Fbar\Fbar},\>
J_{SSFFS},\>
J_{SS\Fbar\Fbar S}.
\label{eq:novectorfunctions}
\eeq
In a second category are 5 functions which involve vector fields, but for which there are no corresponding
$\epsilon$-scalar contributions. This occurs when all Feynman diagram contributions to the function have only
vector lines that terminate (at one end, at least) in a vector-scalar-scalar vertex, since in that case the vector index will be contracted with a momentum, which lives in only $d$ dimensions, not 4, thus projecting out the 
$\epsilon$-scalar components.
Therefore, these 5 functions are again the same in the \DRbarprime and \MSbar schemes:
\beq
H_{FFFVSS},\>
H_{F\Fbar\Fbar VSS},\>
H_{\Fbar\Fbar FVSS},\>
H_{SSSSSV},\>
H_{SSSVVV}.\>
\eeq
The remaining 60 functions do involve $\epsilon$-scalar contributions in at least one contributing diagram, and are therefore different in the
\DRbarprime and \MSbar schemes:
\beq
&& \!\!\!\!\!\!\!
H_{VVSSSS},\>
H_{SSVVSS},\>
H_{VVVSSS},\>
H_{VVSSVS},\>
H_{SSVVVV},\>
H_{SVVVSV},\>
\nonumber \\ && \!\!\!\!\!\!\!
K_{SSSSSV},\>
K_{SSSSVV},\>
K_{SSSVVS},\>
K_{VVSSSS},\>
K_{SSSVVV},\>
K_{VVSSVS},\>
\nonumber \\ && \!\!\!\!\!\!\!
K_{SSVVVV},\>
K_{VVSVVS},\>
J_{SSVSS},\>
J_{SSVVS},\>
G_{VSVVS},\>
H_{\mbox{gauge},S},\>
\nonumber \\ && \!\!\!\!\!\!\!
K_{\mbox{gauge},S},\>
K_{\mbox{gauge},SS},\>
H_{FFVVFF},\>
H_{FFVV\Fbar\Fbar},\>
H_{F\Fbar VVF\Fbar},\>
H_{\Fbar\Fbar VV\Fbar\Fbar},\>
\nonumber \\ && \!\!\!\!\!\!\!
H_{FFFVVV},\>
H_{F\Fbar\Fbar VVV},\>
K_{FFFVVF},\>
K_{FF\Fbar VV\Fbar},\>
K_{\Fbar\Fbar FVVF},\>
K_{\Fbar F\Fbar VVF},\>
\nonumber \\ && \!\!\!\!\!\!\!
K_{\Fbar\Fbar\Fbar VV\Fbar},\>
K_{VVFFFF},\>
K_{VVFF\Fbar\Fbar},\>
K_{VV\Fbar\Fbar\Fbar\Fbar},\>
K_{\mbox{gauge},FF},\>
K_{\mbox{gauge},\Fbar\Fbar},\>
\nonumber \\ && \!\!\!\!\!\!\!
H_{FFSVFF},\>
H_{FFSV\Fbar\Fbar},\>
H_{F\Fbar SV\Fbar F},\>
H_{F\Fbar SVF\Fbar},\>
H_{\Fbar\Fbar SV\Fbar\Fbar},\>
H_{F\Fbar FSVV},\>
\nonumber \\ && \!\!\!\!\!\!\!
H_{FF\Fbar SVV},\>
H_{\Fbar\Fbar\Fbar SVV},\>
K_{FFFSVF},\>
K_{FF\Fbar SV\Fbar},\>
K_{\Fbar\Fbar FSVF},\>
K_{\Fbar F\Fbar SVF},\>
\nonumber \\ &&\!\!\!\!\!\!\!
K_{\Fbar FF SV\Fbar},\>
K_{\Fbar\Fbar\Fbar SV\Fbar},\>
K_{SSSVFF},\>
K_{SSSV\Fbar\Fbar},\>
K_{SSVVFF},\>
K_{SSVV\Fbar\Fbar},\>
\nonumber \\ &&\!\!\!\!\!\!\!
K_{VVSSFF},\>
K_{VVSS\Fbar\Fbar},\>
K_{VVSVFF},\>
K_{VVSV\Fbar\Fbar},\>
H_{\mbox{gauge}},\>
K_{\mbox{gauge}}.
\label{eq:genloopfunctions}
\eeq
The main letters $E$, $G$, $H$, $J$, $K$,$L$,  correspond to the parent Feynman diagram topology, and the subscripts encode the information about the types of propagators in a canonical ordering, as 
shown in Figure \ref{fig:diagramtopologies}, and explained in detail in ref.~\cite{Martin:2017lqn}. The distinction between $F$ and $\overline F$ is that the latter contains a chirality-flipping fermion mass insertion. Note that in many cases involving gauge boson interactions, more than one Feynman diagram contributes to a given function with a fixed structure of gauge invariants. For some of these, the word ``gauge" in a subscript indicates combinations of diagram topologies involving multiple gauge vector boson or ghost propagators with a common group theoretic structure. 
The difference in the present paper is that for each vector propagator, one also includes the corresponding $\epsilon$-scalar contribution in the \DRbarprime function. The results of evaluating all 89 functions appearing in $V^{(3)}$ are given in the ancillary file {\tt functionsDRED.anc}.
These constitute the main new results of this paper. For any given softly broken supersymmetric gauge theory, one can plug in the results for the renormalized field-dependent masses and couplings, as specified above, into eqs.~(3.2)-(3.32) of ref.~\cite{Martin:2017lqn} to evaluate the 3-loop effective potential. 

At 3-loop order, there is a qualitatively new practical problem not encountered at 1-loop and 2-loop orders; the presence of doubled bosonic propagator lines carrying the same momentum (the propagators labeled 1 and 2 in topologies $J$, $K$, and $L$ of Figure \ref{fig:diagramtopologies}) with small or vanishing squared masses can give rise to possible logarithmic infrared singularities. In the case of doubled Goldstone boson propagators, this problem was noted in the context of the 3-loop effective potential in ref.~\cite{Martin:2013gka}. Besides causing infrared divergence problems when the renormalization scale choice leads to small Goldstone boson squared masses, it manifests as imaginary parts of the effective potential at the minima of its real part when the tree-level Goldstone boson squared masses are negative. These imaginary parts are spurious, in the sense that they are not associated with any actual instability of the true vacuum state. This can occur for perfectly reasonable choices of the renormalization scale, including (see ref.~\cite{Martin:2013gka}) in the case of the actual parameters of the Standard Model. In general, this problem can be systematically defeated by resummation, as shown in 
refs.~\cite{Martin:2014bca,Elias-Miro:2014pca}, with further elucidations in refs.~\cite{Pilaftsis:2015bbs,Kumar:2016ltb,Espinosa:2016uaw,Braathen:2016cqe,Braathen:2017izn,Martin:2017lqn}. Doubled propagators of massless gauge bosons can also \cite{Martin:2017lqn} cause infrared divergences in the effective potential; these are benign, in the sense that they are also eliminated in the process of resumming the Goldstone boson contributions. It was also demonstrated in ref.~\cite{Martin:2017lqn} that infrared divergences from doubled massless fermion lines do not occur at 3-loop order. 

Even in cases without infrared divergences due to doubled bosonic propagators, the presence of vanishing squared mass arguments
can cause practical problems, because the 3-loop integral functions for generic squared mass arguments will often
contain individual terms with denominators with powers of the arguments when written in terms of the basis integrals. Although these do not give true infrared singularities, demonstrating this and obtaining expressions suitable for numerical evaluation often requires taking limits of basis integral functions that may not be immediately obvious.

To deal efficiently and systematically with these issues 
in particular cases, it is useful to have expansions of the basis integral functions for small squared mass arguments $\delta$. A complete list of such expansions for every combination of squared mass arguments satisfying
\beq
0 \,<\, \delta \,\ll\, u,v,w,x,y,z
\eeq
is provided in an ancillary file {\tt expdelta.anc} distributed with this paper. (Reference \cite{Martin:2017lqn} provided an ancillary file {\tt expzero.anc} with a subset of these expansions, which was less complete but sufficient for the special cases needed there, namely those encountered in the Standard Model.) These expansions are derived using the differential equations that the basis integrals satisfy (obtained in ref.~\cite{Martin:2016bgz}), and are given to order $\delta^{5}$ for $I$, $F$ and $\overline F$ functions, order $\delta^4$ for $\overline I$ and $G$ functions, and order $\delta^3$ for $K$ and $H$. At 3-loop order, the expansions can contain up to three powers of $\lnbar(\delta)$. 
Whenever a squared mass argument vanishes, or should be treated as small, one can replace it with $\delta$, and then use these expansions to evaluate the leading order contribution to the effective potential as $\delta \rightarrow 0$.
Poles in $\delta$ always cancel, and possible infrared divergences in individual diagrams then manifest themselves as residual powers of $\lnbar(\delta)$, which must also cancel from the minimization conditions for the effective potential, and from associated physically meaningful quantities. This provides a useful check in examples, including the ones mentioned below.

\section{Checks from unbroken supersymmetry\label{sec:unbrokenSUSY}}
\setcounter{equation}{0}
\setcounter{figure}{0}
\setcounter{table}{0} 
\setcounter{footnote}{1}

Consider a supersymmetric theory, with no explicit supersymmetry breaking terms. (For reviews of supersymmetry using notations and conventions consistent with the following, see \cite{Martin:1997ns,Dreiner:2023yus}.)
As shown by Zumino in ref.~\cite{Zumino:1974bg}, at a supersymmetric minimum of the tree-level potential, the full effective potential must vanish at each order in perturbation theory. In the case of non-gauge theories like the Wess-Zumino model, this was used in ref.~\cite{Martin:2017lqn} as a check on the 24 contributions at 3-loop order in eq.~(\ref{eq:novectorfunctions}) above, which do not involve vector bosons. I have now extended these checks to various supersymmetric gauge theory special cases.

For an example that is simple enough to analyze explicitly in text, consider a supersymmetric $U(1)$ gauge theory with gauge coupling $g$ and two chiral superfields $\Phi_+$ and $\Phi_-$ with charges $+1$ and $-1$ respectively,
and a superpotential mass term
\beq
W &=& \mu \Phi_+ \Phi_- .
\eeq
The corresponding complex scalar fields can be written in terms of canonically normalized real components as
\beq
\phi_+ = \frac{1}{\sqrt{2}} (R_1 + i R_2),\qquad
\phi_- = \frac{1}{\sqrt{2}} (R_3 + i R_4).
\eeq
There are three Weyl fermions, $\psi_1 = \psi_+$, $\psi_2 = \psi_-$, and $\psi_3 = \lambda$, the latter being the
gaugino field. Finally, there is a single vector boson, $A^\mu$. In order to main unbroken supersymmetry, the 
background scalar field components for $\Phi_+$ and $\Phi_-$ have been chosen to both vanish, leading to
a tree-level potential $V^{(0)} = 0$. Then the 
gaugino and gauge boson masses vanish, while the 
chiral fermion and scalar squared masses are all equal:
\beq
M_3^2 = m_A^2 &=& 0 ,
\\
M_1^2 = M_2^2 = m_1^2 = m_2^2 = m_3^2 = m_4^2 &=& z,
\eeq
with
\beq
M_{12} = M_{21} = \sqrt{z} \equiv \mu
.
\eeq
There are no scalar cubic interactions, and the non-vanishing quartic scalar interactions are:
\beq
\lambda^{1111} = \lambda^{2222} = \lambda^{3333} = \lambda^{4444} &=& 3 g^2,
\\
\lambda^{1122} = \lambda^{3344} &=& g^2, 
\\
\lambda^{1133} = \lambda^{1144} = \lambda^{2233}  = \lambda^{2244} &=& -g^2,
\eeq
and permutations thereof,
while the non-vanishing Yukawa couplings, stemming from gaugino interactions with scalar and chiral fermion pairs, are
\beq
Y^{113} = i Y^{213} = -Y^{323} = -i Y^{423} = g,
\eeq
and equal values when the last two (fermion) indices are interchanged.
The non-vanishing vector-scalar-scalar couplings are
\beq
g^{A21} = -g^{A12} = g^{A34} = -g^{A43} = g, 
\eeq
and the vector-fermion-fermion couplings are
\beq
g^{A1}_1 = -g^{A2}_2 = g.
\eeq
There are no vector-vector-scalar interactions (because the gauge symmetry is not spontaneously broken) and no vector-vector-vector interactions (because the gauge symmetry is Abelian).
The 1-loop part of the effective potential, evaluated from eq.~(\ref{eq:oneloopV}), is
\beq
V^{(1)} = 4 f(z) - 4 f(z) = 0,
\eeq 
where the two terms come from the scalar and fermion contributions to the supertrace, respectively, and there is no contribution from the massless vectors because $f(0) = 0$. The 2-loop contribution from eq.~(\ref{eq:V2loopgeneral}) is
\beq
V^{(2)} &=& g^2 \left [f_{SS}(z,z) + 4 f_{FFS}(0,z,z) + f_{VSS}(0,z,z) + f_{FFV}(z,z,0) - z f_{\overline F\overline F V}(z,z,0) \right ].
\phantom{xxx}
\label{eq:SQEDV2}
\eeq
This also vanishes, due to non-trivial cancellations between these functions, obtained by plugging in their expressions
in terms of the basis integrals from the file {\tt functionsDRED.anc}. Finally, the three-loop contribution obtained from the general form specified in eqs.~(3.2)-(3.32) of ref.~\cite{Martin:2017lqn} is
\beq
V^{(3)} &=&
g^4 \Bigl [
L_{SSSS}(z,z,z,z) + \tfrac{3}{2} E_{SSSS}(z,z,z,z) 
+ 4 z H_{F\Fbar SS F \Fbar}(0,z,z,z,0,z)
\nonumber \\ &&
+ 4 J_{SSFFS}(z,z,0,z,z) 
+ 4 K_{FFFSSF}(z,z,0,z,z,0)
+ 4 z K_{\Fbar\Fbar FSSF}(z,z,0,z,z,0)
\nonumber \\ &&
+ 8 K_{FFFSSF}(0,0,z,z,z,z) + 4 K_{SSFFFF}(z,z,0,z,0,z)
-\tfrac{1}{2} H_{SSVVSS}(z, z, 0, 0, z, z) 
\nonumber \\ &&
- 2 J_{SSVSS}(z,z,0,z,z)
+ K_{SSSVVS}(z,z,z,0,0,z) 
+ K_{VVSSSS}(0,0,z,z,z,z)
\nonumber \\ &&
+\tfrac{1}{2} z^2 H_{\Fbar \Fbar VV\Fbar \Fbar }(z, z, 0, 0, z, z)
+ z H_{F\Fbar VVF\Fbar }(z, z, 0, 0, z, z) 
- 2 z H_{FFVV\Fbar \Fbar }(z, z, 0, 0, z, z) 
\nonumber \\ &&
+ \tfrac{1}{2} H_{FFVVFF}(z, z, 0, 0, z, z) 
+ z^2 K_{\Fbar \Fbar \Fbar VV\Fbar }(z, z, z, 0, 0, z) 
+ z K_{\Fbar \Fbar FVVF}(z, z, z, 0, 0, z) 
\nonumber \\ &&
- 4 z K_{\Fbar F\Fbar VVF}(z, z, z, 0, 0, z) 
+ z K_{FF\Fbar VV\Fbar }(z, z, z, 0, 0, z) 
+ K_{FFFVVF}(z, z, z, 0, 0, z) 
\nonumber \\ &&
+ z^2 K_{VV\Fbar \Fbar \Fbar \Fbar }(0, 0, z, z, z, z) 
- 2 z K_{VVFF\Fbar \Fbar }(0, 0, z, z, z, z)
+ K_{VVFFFF}(0, 0, z, z, z, z)   
\nonumber \\ &&
+ 4 z H_{\Fbar \Fbar FVSS}(z, z, 0, 0, z, z) 
+ 4 H_{FFFVSS}(z, z, 0, 0, z, z) 
+ 4 z K_{\Fbar \Fbar FSVF}(z, z, 0, z, 0, z)
\nonumber \\ &&
- 8 z K_{\Fbar FFSV\Fbar }(z, z, 0, z, 0, z) 
+ 4 K_{FFFSVF}(z, z, 0, z, 0, z) 
- 4 K_{SSSVFF}(z, z, z, 0, 0, z) 
\nonumber \\ &&
+ 2 z K_{VVSS\Fbar \Fbar }(0, 0, z, z, z, z) 
- 2 K_{VVSSFF}(0, 0, z, z, z, z)
\Bigr ] .
\label{eq:SQEDV3}
\eeq
As noted at the end of the previous section, to evaluate this properly 
one may first change the 0 arguments to $\delta$,
then after using the results in {\tt functionsDRED.anc}, apply the expansions in {\tt expdelta.anc} to keep only
non-vanishing terms as $\delta \rightarrow 0$. Most of the functions in
eq.~(\ref{eq:SQEDV3}) are individually completely smooth in the limit $\delta \rightarrow 0$. The exceptions, which have only simple logarithmic singularities, are:
\beq
K_{VV\Fbar \Fbar \Fbar \Fbar }(\delta, \delta, z, z, z, z)
&=& 12\, \lnbar(\delta) [z + A(z)]^2/z^2 + \ldots,
\\
K_{VV\Fbar \Fbar FF}(\delta, \delta, z, z, z, z)
&=& 12\, \lnbar(\delta) [z + A(z)]^2/z + \ldots
\\
K_{VVFFFF}(\delta, \delta, z, z, z, z)
&=& 12\, \lnbar(\delta) [z + A(z)]^2 + \ldots,
\eeq
where the ellipses represent terms that are finite as $\delta \rightarrow 0$.
Since these functions appear in eq.~(\ref{eq:SQEDV3}) with coefficients proportional to $z^2$, $-2z$, and $1$ respectively, the $\lnbar(\delta)$ terms are seen to successfully cancel in the complete expression. In fact, the whole expression for $V^{(3)}$ vanishes in the
limit $\delta \rightarrow 0$, as required, due to non-trivial cancellations between the various functions. This becomes apparent after expressing the results in terms of the renormalized 3-loop basis integrals.

The simple example above does not come close to completely testing the results obtained in {\tt functionsDRED.anc}, because of the absence of scalar-scalar-scalar, vector-vector-scalar, and vector-vector-vector interactions, and the absence of superpotential Yukawa couplings. I have carried out more detailed tests, each including many more terms, as follows: 
\begin{itemize}
\item Supersymmetric $U(1)$ gauge theory with three chiral superfields $\Phi$, $\overline \Phi$, and $\Phi_0$, with charges $+1$, $-1$, and $0$ respectively, and a superpotential $W = y \Phi_0 \Phi \overline\Phi + \mu \Phi \overline\Phi + \frac{1}{2} \mu_0 \Phi_0^2$. Supersymmetry is unbroken when the scalar background fields are taken to vanish.
\item Supersymmetric $U(1)$ gauge theory with two chiral superfields $\Phi$ and $\overline \Phi$, with charges $+1$, $-1$. There is no superpotential. The gauge symmetry is spontaneously broken by equal magnitude
background fields for the scalars, $\overline \varphi = \varphi$. This is a $D$-flat direction, leaving supersymmetry unbroken. 
\item Supersymmetric $SU(n)$ gauge theory with $n=2,3$, with chiral superfields $\Phi_j$ and $\overline \Phi^j$ in the fundamental and anti-fundamental
representations with $j=1,\ldots,n$, and a singlet chiral superfield $\Phi_0$. The superpotential is $W = y \Phi_0 \Phi \overline\Phi + \mu \Phi \overline\Phi + \frac{1}{2} \mu_0 \Phi_0^2$. The background scalar fields are taken to vanish, leaving the gauge symmetry unbroken and maintaining unbroken supersymmetry.  
\item Supersymmetric $SU(n)$ gauge theory with $n=2,3$, with chiral superfields $\Phi_j$ and $\overline \Phi^j$ in the fundamental and anti-fundamental representations, with no superpotential. The scalar fields obtain background values with equal magnitudes along a $D$-flat direction $\Phi_j = \overline \Phi^j = \varphi \delta_{j1}$, breaking the gauge symmetry but again maintaining unbroken supersymmetry.  
\end{itemize}
In each of these cases, I have checked that $V^{(1)} = V^{(2)} = V^{(3)} = 0$ as required by unbroken supersymmetry at tree level. These are highly non-trivial consistency checks on the results obtained in {\tt functionsDRED.anc}, relying on intricate cancellations between the individual contributions after writing them in terms of the renormalized basis integrals. (As one might expect, the cancellations of the individual contributions would not occur if one used the \MSbar functions instead of the correct \DRbarprime ones.)
These cancellations include terms proportional to the infrared regulator $\lnbar(\delta)$ in the 3-loop part,
corresponding to massless vectors and massless scalars along flat directions.

\section{Checks from renormalization group invariance\label{sec:RGinvariance}}
\setcounter{equation}{0}
\setcounter{figure}{0}
\setcounter{table}{0} 
\setcounter{footnote}{1}

Another class of checks, applicable for cases of  non-supersymmetric vacua and softly broken supersymmetric gauge theories, comes from renormalization group invariance. The invariance of the effective potential with respect to changes in the arbitrary renormalization scale $Q$ can be expressed as
\beq
0 = Q \frac{dV_{\rm eff}}{dQ} = 
\left (Q \frac{\partial}{\partial Q} + \sum_{X} \beta_X \frac{\partial}{\partial X} \right ) V_{\rm eff},
\eeq
where $X$ runs over all of the independent \DRbarprime parameters of the theory, including the background scalar field(s) $\varphi$, the masses and couplings that may depend on the $\varphi$, and a field-independent contribution to the tree-level potential, which I will denote below by $\Lambda$. The beta functions for the parameters $X$ are given in a loop expansion by
\beq
\beta_X &=& 
\frac{1}{16\pi^2} \beta_X^{(1)} + 
\frac{1}{(16\pi^2)^2} \beta_X^{(2)} + 
\frac{1}{(16\pi^2)^3} \beta_X^{(3)} + 
\ldots,
\eeq
and in the particular case of the background scalar fields one writes
$
\beta_{\varphi} = -\gamma^S \varphi ,
$
where $\gamma^S$ is the scalar anomalous dimension, not to be confused with the chiral superfield anomalous dimension.
Therefore, at each loop order $\ell$, consistency requires
\beq
Q \frac{\partial}{\partial Q} V^{(\ell)} + \sum_{n=0}^{\ell-1} \left ( \sum_X \beta_X^{(\ell-n)} \frac{\partial}{\partial X} V^{(n)} \right ) &=& 0.
\label{eq:RGcheck}
\eeq
To evaluate the first term in eq.~(\ref{eq:RGcheck}), the derivatives with respect to $Q$ of the basis integrals, and of the 9 two-loop functions and the 89 three-loop functions, are given for convenience in an ancillary file 
{\tt QdQDRED.anc} distributed with this paper. Since most of the $\beta_X$ functions are known from previous work, evaluating eq.~(\ref{eq:RGcheck}) for each $\ell$ in particular cases in principle gives non-trivial checks on the results of the present paper in the file {\tt functionsDRED.anc}. 

However, there are two missing pieces of information. First, although the 2-loop and 3-loop contributions to the anomalous dimensions of the chiral superfields were calculated in refs.~\cite{West:1984dg,Jack:1996qq}, the anomalous dimensions of the scalar components are different, and were only previously known to 2-loop order. Second, the beta function of the field-independent vacuum energy $\Lambda$ was only previously known at 2-loop order. Therefore, by demanding that eq.~(\ref{eq:RGcheck}) holds for $\ell=1,2,3$ in a variety of cases, I have been able to derive and then check these missing results. I will first provide these results, and then briefly review the list of special case models used to infer and check them.

Consider a supersymmetric gauge theory with chiral superfields $\Phi_i$, and a superpotential
\beq
W &=& \frac{1}{6} y^{ijk}  \Phi_i \Phi_j \Phi_k + \frac{1}{2} \mu^{ij} \Phi_i \Phi_j 
\eeq
involving Yukawa couplings $y^{ijk}$ and supersymmetric masses $\mu^{ij}$, and soft supersymmetry breaking terms 
\beq
-{\cal L}_{\rm soft}
&=&
\left ( \frac{1}{6} a^{ijk} \phi_i \phi_j \phi_k + \frac{1}{2} b^{ij} \phi_i \phi_j  + \frac{1}{2} M_a \lambda^a \lambda^a \right ) + {\rm c.c.} +
(m^2)_j^i \phi_i \phi^{*j}
+ \Lambda,
\eeq
where $\phi_i$ are the scalar components of $\Phi_i$, and $\lambda^a$ are the gaugino fields. Here $a^{ijk}$ and $b^{ij}$ are holomorphic scalar cubic and scalar squared mass terms respectively, $M_a$ are the gaugino masses, and $(m^2)_j^i$ are
the non-holomorphic scalar squared masses.
The last term, the field-independent vacuum energy $\Lambda$, is irrelevant to the (non-gravitational) dynamics of the theory and therefore generally omitted, but its presence is necessary to maintain renormalization-scale invariance of $V_{\rm eff}$. Note that in all checks below I have assumed that there are no tadpole couplings in $W$ or $-{\cal L}_{\rm soft}$ associated with gauge-singlet chiral superfields. I also assume that there is at most one $U(1)$ component in the gauge group, to avoid the complication of kinetic mixing between different Abelian gauge fields. Both of these assumptions hold in the case of the minimal supersymmetric Standard Model (MSSM).

The gauge group is assumed to have couplings $g_a$, with generators $({\bf t}^a)_i{}^j$. The notation for group theory invariants will closely follow that of the review in Chapter 11 of ref.~\cite{Dreiner:2023yus}. For each distinct group component, the dimension (number of Lie algebra generators) and the quadratic Casimir invariant are denoted $d_a$ and $G_a$ respectively. The quadratic Casimir invariant of an irreducible representation carrying a flavor index $i$ is denoted
$C_a(i)$, where
\beq
({\bf t}^a {\bf t}^a)_i{}^j &=& C_a(i) \delta_i^j.
\eeq
For an irreducible representation $r$, the Dynkin index is $T_a(r)$, defined by
\beq
{\rm Tr}_r [{\bf t}^a {\bf t}^b] &=& \delta_{ab} T_a(r),
\eeq
and the sum of the $T_a(r)$ over all of the chiral supermultiplet representations is
\beq
S_a &=& \sum_r T_a(r),
\eeq
Similarly, define 
\beq
S_{ab} &=& \sum_r T_a(r) C_b(r),
\\
S_{abc} &=& \sum_r T_a(r) C_b(r) C_c(r).
\eeq
For example, for a supersymmetric $U(1)$ gauge theory with chiral superfields $\Phi_i$ with charges $q_i$, one
has $d_a=1$, $G_a = 0$, $C_a(i) = q_i^2$, $S_a = \sum_i q_i^2$, $S_{aa} = \sum_i q_i^4$, and $S_{aaa} = \sum_i q_i^6$.
For a supersymmetric $SU(n_c)$ gauge theory with $n_f$ flavors of fundamental and anti-fundamental chiral superfields,
one has $d_a = n_c^2 - 1$, $G_a = n_c$, $C_a(i) = (n_c^2 - 1)/(2 n_c) \equiv C_f$ for each $i$, and $S_a = n_f$, 
$S_{aa} = n_f C_f$, and $S_{aaa} = n_f C_f^2$.

The DRED beta functions for the gauge couplings $g_a$ were found at 2-loop order in \cite{Jones:1974pg,Jones:1983vk}, and at 3-loop order in \cite{Jack:1996vg} by making use of results in \cite{Novikov:1985rd,Shifman:1986zi}. Using the notations above, they are:
\beq
\beta_{g_a}^{(1)} &=& g_a^3 \left ( S_a - 3 G_a \right ),
\label{eq:betagaoneloop}
\\[3pt]
\beta_{g_a}^{(2)} &=& 2 g_a^5 G_a \left ( S_a -3 G_a \right )
+ 4 g_a^3 g_b^2 S_{ab}
- g_a^3 y^{ijk} y_{ijk} C_a(i)/d_a ,
\\[3pt]
\beta_{g_a}^{(3)} &=& g_a^7 G_a \left (S_a - 3 G_a \right )
\left (7 G_a - S_a \right )
+ 8 g_a^5 g_b^2 G_a S_{ab}
+ 6 g_a^3 g_b^4 S_{ab} \left (3 G_b -  S_b \right )
\nonumber \\[3pt] &&
-\, 8 g_a^3 g_b^2 g_c^2 S_{abc}
- 2 g_a^5 y^{ijk} y_{ijk} C_a(i) G_a/d_a
+\, g_a^3 g_b^2 y^{ijk} y_{ijk}
\left [ C_b(i) - 6  C_b(j) \right] C_a(i)/d_a
\phantom{xxx}
\nonumber \\[3pt] &&
+\, g_a^3 y^{ijk} y_{ijl} y_{kmn} y^{lmn}
\left [3 C_a(i)/2 + C_a(k)/4 \right ]/d_a .
\eeq
The anomalous dimension of the chiral superfield $\Phi_i$, and the anomalous dimension of its scalar component $\phi_i$, have the same general form:
\beq
\gamma_i^{(1)j} &=& \tfrac{1}{2} y_{ikl} y^{jkl} + n_1 \delta_i^j g_a^2 C_a(i)
,
\label{eq:gammageneralone}
\\
\gamma_i^{(2)j} &=&
-\tfrac{1}{2} y_{ikl} y^{jkm} y^{lnp} y_{mnp}
+ g_a^2 y_{ikl} y^{jkl} [n_2 C_a(k) + n_3 C_a(i)]
\nonumber \\ &&
+\, \delta_i^j g_a^2 C_a(i)
\left [ n_4 g_a^2 S_a + n_5 g_b^2 C_b(i) + n_6 g_a^2 G_a \right ]
,
\label{eq:gammageneraltwo}
\\
\gamma_i^{(3)j} &=&
- \tfrac{1}{8} y_{ikl} y^{jpq} y^{kmn} y_{pmn} y^{lrs} y_{qrs}
- \tfrac{1}{4} y_{ikl} y^{jkm} y^{lnp} y_{snp} y^{sqr} y_{mqr}
\nonumber \\ &&
+\, y_{ikl} y^{jkm} y^{lnp} y_{mnq} y^{qrs} y_{prs}
+ \tfrac{3}{2}\zeta_3\, y_{ikl} y^{jpq} y^{kmn} y^{lrs} y_{pmr} y_{qns}
\nonumber \\ &&
+\, g_a^2 y_{ikl} y^{jkm} y^{lnp} y_{mnp}
\bigl [
  n_{7} C_a(i)
+ n_{8} C_a(k)
+ n_{9} C_a(l) 
+ n_{10} C_a(p)
\bigr ]
\nonumber \\ &&
+\, g_a^2 g_b^2 y_{ikl} y^{jkl}
\bigl [
n_{11} C_a(i) C_b(i)
+ n_{12} C_a(i) C_b(k)
+\, n_{13} C_a(k) C_b(k)
+ n_{14} C_a(k) C_b(l)
\bigr ]
\nonumber \\ &&
+\, g_a^4  y_{ikl} y^{jkl} \bigl [n_{15} S_a C_a(i) + n_{16} S_a C_a(k)
+ n_{17} G_a C_a(i) + n_{18} G_a C_a(k) \bigr ] 
\nonumber \\ &&
+\, \delta_i^j g_a^2 C_a(i) \Bigl \{
g_a^4 \bigl[ n_{19} S_a^2 + n_{20} S_a G_a + n_{21} G_a^2
\bigr ]
\nonumber \\ &&
\qquad\qquad 
+\, g_a^2 g_b^2 \bigl[
n_{22} S_a C_b(i) + n_{23} G_a C_b(i) + n_{24} S_{ab}
\bigr]
\nonumber \\ &&
\qquad\qquad +\, n_{25} g_b^2 g_c^2 C_b(i) C_c(i)
+ n_{26} g_a^2 y^{klm} y_{klm} C_a(k)/d_a
\Bigr\}
.
\label{eq:gammageneralthree}
\eeq
However, as is well-known, some of the coefficients $n_1, \ldots, n_{26}$ of the gauge-coupling-dependent terms differ for the chiral superfield and its scalar component. Indeed, the coefficients for the scalar component are dependent on the choice of gauge-fixing, while the coefficients for the chiral superfield are not.
For the chiral superfield anomalous dimension, the results are \cite{West:1984dg,Jack:1996qq}:
\beq
&&
n_1 = -2,\qquad
n_2 = 2,\qquad
n_3 = -1,\qquad
n_4 = 2,\qquad
n_5 = 4,\qquad
n_6 = -6,
\nonumber
\\
&&
n_{7} = 2 + 3 \zeta_3,\qquad
n_{8} = 1 - 3 \zeta_3,\qquad
n_{9} = 4 - 6 \zeta_3,\qquad
n_{10} = -4 + 6 \zeta_3,\qquad
\nonumber
\\
&&
n_{11} = 4 - 15 \zeta_3,\qquad
n_{12} = -8 + 12 \zeta_3,\qquad
n_{13} = -12 - 6 \zeta_3,\qquad
n_{14} = -2 + 18 \zeta_3,\qquad
\nonumber
\\
&&
n_{15} = -1,\qquad
n_{16} = -4,\qquad
n_{17} = 3 - 3 \zeta_3,\qquad
n_{18} = 12 - 6 \zeta_3,\qquad
n_{19} = 2,\qquad
\nonumber
\\
&&
n_{20} = -2 + 24 \zeta_3,\qquad
n_{21} = -12,\qquad
n_{22} = -4,\qquad
n_{23} = 12,\qquad
\nonumber
\\
&&
n_{24} = 20 - 24 \zeta_3,\qquad
n_{25} = -16,\qquad
n_{26} = -5.
\eeq
For the scalar component, I find that the Landau gauge coefficients are instead:
\beq
&&
n_1 = -1,\qquad
n_2 = 2,\qquad
n_3 = -1,\qquad
n_4 = 1,\qquad
n_5 = 2,\qquad
n_6 = -9/4,
\nonumber
\\
&&
n_{7} = 2 + 3 \zeta_3,\qquad
n_{8} = 1 - 3 \zeta_3,\qquad
n_{9} = 4 - 6 \zeta_3,\qquad
n_{10} = -4 + 6 \zeta_3,\qquad
\nonumber
\\
&&
n_{11} = 4 - 12 \zeta_3,\qquad
n_{12} = -8 + 12 \zeta_3,\qquad
n_{13} = -12,\qquad
n_{14} = -2 + 12 \zeta_3,\qquad
\nonumber
\\
&&
n_{15} = -1,\qquad
n_{16} = -4,\qquad
n_{17} = 7/2 - 9 \zeta_3/2,\qquad
n_{18} = 12 - 6 \zeta_3,\qquad
n_{19} = 1,\qquad
\nonumber
\\
&&
n_{20} = -9/4 + 12 \zeta_3,\qquad
n_{21} = -13/16 - 63 \zeta_3/8,\qquad
n_{22} = -2,\qquad
n_{23} = -1 + 15 \zeta_3,\qquad
\nonumber
\\
&&
n_{24} = 10 - 12 \zeta_3,\qquad
n_{25} = -12,\qquad
n_{26} = -5/2.
\label{eq:coeffsgammaS}
\eeq
The first 6 of these are not new, having been obtained in ref.~\cite{Martin:2001vx} from the \DRbarprime 2-loop effective potential.

The different roles played by the chiral superfield anomalous dimension and the scalar component field anomalous dimension are as follows. The former enters into the beta functions for superpotential parameters, according to
\beq
\beta_{y^{ijk}} &=& \gamma_n^i y^{njk} + (i \leftrightarrow j) + (i \leftrightarrow k)
,
\\
\beta_{\mu^{ij}} &=& \gamma_n^i \mu^{nj} +  (i \leftrightarrow j)
,
\eeq
valid at all orders in perturbation theory.
The scalar component field anomalous dimension $\gamma^S$ is instead related to the beta function of the background scalar fields $\varphi_i$, according to
\beq
\beta_{\varphi_i} &=& -(\gamma^S)_i^j \varphi_j,
\eeq
for use with $X = \varphi_i$ in eq.~(\ref{eq:RGcheck}).

For the sake of completeness, I also review the beta functions for the soft supersymmetry-breaking parameters as needed below, again following closely the notation of the review in Chapter 11 of ref.~\cite{Dreiner:2023yus}. This can be done most efficiently in terms of differential operators in coupling-constant space that act on the chiral superfield anomalous dimensions:
\beq
\Omega &=& \frac{1}{2} M_a g_a \frac{\partial}{\partial g_a} - a^{ijk} \frac{\partial}{\partial y^{ijk}},
\\
\Omega^* &=& \frac{1}{2} M_a^* g_a \frac{\partial}{\partial g_a} - a_{ijk} \frac{\partial}{\partial y_{ijk}}.
\eeq
The beta functions for $M_a$, $a^{ijk}$, and $b^{ij}$ were found at 2-loop order results in 
refs.~\cite{Martin:1993yx,Martin:1993zk,Yamada:1994id,Jack:1994kd}, and extended by
refs.~\cite{Jack:1997pa,Jack:1997eh} to all orders in perturbation theory, 
\beq
\beta_{M_a} &=& 2 \Omega (\beta_{g_a}/g_a)
,
\\[3pt]
\beta_{a^{ijk}} &=& \left [
\gamma^i_n a^{njk} - 2 y^{njk} \Omega(\gamma_n^i) \right ]
+ (i \leftrightarrow j)
+ (i \leftrightarrow k), 
\label{eq:betaaijk}
\\[3pt]
\beta_{b^{ij}} &=& \left [
\gamma^i_n b^{nj} - 2 \mu^{nj} \Omega(\gamma_n^i) \right ]
+ (i \leftrightarrow j).
\eeq
using spurion methods as proposed in Ref.~\cite{Yamada:1994id}. However, in cases with gauge-singlet chiral superfields, $\beta_{b^{ij}}$ contains extra terms not captured by the above. The results were given in refs.~\cite{Yamada:1994id,Martin:1993zk,Jack:1994kd} at two-loop order: 
\beq
\beta^{(1)}_{b^{ij}} & = &
\frac{1}{2} b^{il} y_{lmn} y^{mnj} + \frac{1}{2} y^{ijl} y_{lmn} b^{mn}
+ \mu^{il} y_{lmn} a^{mnj}
- 2 \left (b^{ij} - 2 M_a \mu^{ij} \right ) g_a^2 C_a (i)  
\nonumber \\ && 
+ (i \leftrightarrow j),
\\
\beta^{(2)}_{b^{ij}} & = &
-\frac{1}{2} b^{il} y_{lmn} y^{pqn} y_{pqr} y^{mrj}
-\frac{1}{2} y^{ijl} y_{lmn} b^{mr} y_{pqr} y^{pqn} 
-\frac{1}{2} y^{ijl} y_{lmn} \mu^{mr} y_{pqr} a^{pqn}
\nonumber \\ &&
- \mu^{il} y_{lmn} a^{npq} y_{pqr}  y^{mrj} 
- \mu^{il} y_{lmn} y^{npq} y_{pqr} a^{mrj}
+ 2 y^{ijl} y_{lpq} \left ( b^{pq} - \mu^{pq} M_a \right ) g_a^2 C_a(p)
\nonumber \\ &&
+ \left ( b^{il} y_{lpq} y^{pqj} + 2 \mu^{il} y_{lpq} a^{pqj}
- 2 \mu^{il} y_{lpq} y^{pqj} M_a \right )
g_a^2 \left[ 2 C_a(p) - C_a(i) \right ] 
\nonumber \\ &&
+  \left ( 2 b^{ij} - 8 \mu^{ij} M_a \right )
g_a^2 C_a(i) \left [g_a^2 S_a + 2 g_b^2 C_b(i) - 3 g_a^2 G_a \right ] 
\nonumber \\ &&
+ (i \leftrightarrow j).
\eeq
This will be sufficient for the examples considered below. A way of finding $\beta_{b^{ij}}$ at arbitrary loop order in terms of the chiral superfield anomalous dimension is given in ref.~\cite{Jack:2001ew}. For the non-holomorphic soft squared masses $(m^2)_i^j$, the result is \cite{Jack:1998iy}:
\beq
\beta_{(m^2)_i^j} &=&  g_a^2 (\boldsymbol{t^a})_i{}^j A_{a}+
\biggl \{
2 \Omega \Omega^*
+ \left ( |M_a|^2 + X_a \right ) g_a \frac{\partial}{\partial g_a}
\nonumber \\ && \quad\qquad
+ \left [(m^2)_k^n y^{kpq} + (m^2)_k^p y^{nkq} + (m^2)_k^q y^{npk} \right ]
\frac{\partial}{\partial y^{npq}}
\nonumber \\ && \quad\qquad
+ \left [(m^2)^k_n y_{kpq} + (m^2)^k_p y_{nkq} + (m^2)^k_q y_{npk} \right ]
\frac{\partial}{\partial y_{npq}}
\biggr \}\, \gamma_i^j\,,
\label{eq:betam2ijgeneral}
\eeq
with
\beq
X_a &=&
\frac{1}{16\pi^2} X_a^{(1)}
+ \frac{1}{(16\pi^2)^2} X_a^{(2)}
+ \dotsb
,
\\
A_a &=&
\frac{1}{16\pi^2} A_a^{(1)}
+ \frac{1}{(16\pi^2)^2} A_a^{(2)}
+ \frac{1}{(16\pi^2)^3} A_a^{(3)}
+ \dotsb
,
\eeq
where the results needed for 3-loop order $\beta_{(m^2)_i^j}$ are
\cite{Jack:1998iy,Jack:1998uj}:
\beq
X_a^{(1)} &=& 2 g_a^2 \left [ G_a |M_a|^2 - (m^2)_k^k C_a(k)/d_a \right ]
,
\\[3pt]
X_a^{(2)} &=& g_a^4 (10 G_a - 2 S_a) G_a |M_a|^2
- 4 g_a^4 G_a (m^2)_k^k C_a(k)/d_a
- 4 g_a^2 g_b^2 S_{ab} |M_b|^2
\nonumber \\[2pt] &&
+\, g_a^2 y^{kpq} y_{npq} (m^2)_k^n \left [ C_a(p) + \tfrac{1}{2} C_a(k) \right ]/d_a
+ g_a^2 a^{kpq} a_{kpq} C_a(k)/2 d_a.
\phantom{xxxxx..}
\label{eq:Xa2}
\eeq
and the special contributions from Abelian group factors are \cite{Jack:2000jr}:
\beq
A_a^{(1)} &=& 2 (\boldsymbol{t^a})_k{}^l (m^2)_l^k
,
\\[3pt]
A_a^{(2)} &=& (\boldsymbol{t^a})_k{}^l \left [
8 g_b^2 C_b(k) (m^2)_l^k
- 2 (m^2)_n^k y^{npq} y_{lpq} \right ]
,
\\[3pt]
A_a^{(3)} &=& (\boldsymbol{t^a})_k{}^l \Bigl \{
3 (m^2)_l^n y^{kpq} y_{npr} y^{rst} y_{qst}
-\tfrac{3}{2} (m^2)_l^n y^{kpq} y_{pqr} y^{rst} y_{nst}
-\,4 y^{knp} y_{pqr} y^{rst} y_{lst} (m^2)_n^q
\nonumber \\[3pt]
&&
-2 a^{knp} a_{npq} y_{lrs} y^{qrs}
-\tfrac{5}{2} a^{knp} a_{lrs} y^{qrs} y_{npq}
+\, 16 g_b^2 C_b(k) y^{knp} y_{lnp} |M_b|^2
\nonumber \\[3pt]
&&
+ (8 - 24 \zeta_3) g_b^2 C_b(k) a^{knp} a_{lnp}
+\, (12 \zeta_3 - 10) g_b^2 C_b(k) \left [
  a^{knp} y_{lnp} M_b^* + y^{knp} a_{lnp} M_b \right ]
\nonumber \\[3pt]
&&
+\, g_b^2 (m^2)_l^n y^{kpq} y_{npq} [(10 - 24 \zeta_3) C_b(k) -12 C_b(p)]
+\, (16 - 48 \zeta_3) g_b^2 C_b(k) y^{knp} y_{lrp} (m^2)_n^r
\nonumber \\[3pt]
&&
-16 g_b^2 g_c^2 C_b(k) C_c(k) (m^2)_l^k
+\, g_b^2 g_c^2 C_b(k) C_c(k) [ (96 \zeta_3 - 64) |M_b|^2 
\nonumber \\[3pt]
&&
+ (48 \zeta_3 - 40) M_b M_c^* ] \delta^k_l
+\, 12 g_b^4 C_b(k) (3 G_b - S_b) (m^2)_l^k
\Bigr \} .
\label{eq:Aa3}
\eeq
Finally, renormalization group invariance of the effective potential requires non-trivial running
of the field-independent vacuum energy, $X=\Lambda$ in eq.~(\ref{eq:RGcheck}).
The 1-loop and 2-loop contributions were found in ref.~\cite{Martin:2001vx} from the 2-loop \DRbarprime effective potential:
\beq
\beta_\Lambda^{(1)} &=& 
(m^2)_i^j (m^2)_j^i + 2 (m^2)_i^j \mu^{ik} \mu_{jk}  + b^{ij} b_{ij}  - d_a |M_a|^4,
\\[6pt]
\beta_\Lambda^{(2)} &=& 
-y^{ijk} y_{ijl} \left [
(m^2)_k^m (m^2)_m^l + (m^2)_k^m \mu_{mn} \mu^{ln} +  \mu_{kn} \mu^{mn} (m^2)_m^l +
\mu_{km} (m^2)_n^m \mu^{nl} + b^{ml} b_{km}  \right ]
\nonumber \\ &&
-a^{ijk} a_{ijl} \left [(m^2)_k^l +  \mu_{km} \mu^{lm} \right ]
- 2 y^{ijk} y_{ilm} (m^2)_j^l \mu_{kn} \mu^{mn}
- y^{ijk} a_{ijl} \mu_{km} b^{ml}
- y_{ijk} a^{ijl} \mu^{km} b_{ml}
\phantom{xxx}
\nonumber \\ &&
+ 4 g_a^2 C_a(i) \left [(m^2)_i^j (m^2)_j^i + b^{ij} b_{ij} - M_a \mu^{ij} b_{ij} - M^*_a \mu_{ij} b^{ij}
+ 2 (m^2)_i^j \mu^{ik} \mu_{jk}  + 2 \mu^{ij} \mu_{ij} |M_a|^2  \right ]  
\nonumber \\ &&
+ g_a^2 d_a \left (4 S_a - 8 G_a \right ) |M_a|^4
.
\eeq
From the special case examples described below, I was able to deduce the general three-loop result, which is divided into parts
with 0, 2, and 4 powers of gauge couplings:
\beq
\beta_\Lambda^{(3)} &=& \beta_\Lambda^{(3a)} +  \beta_\Lambda^{(3b)} +  \beta_\Lambda^{(3c)}
,
\label{eq:betaL3}
\eeq
where
\beq
&&\!\!\!\beta_\Lambda^{(3a)} \,=\, 
2 y^{ilm} y_{kln} y^{npq} y_{mpq} \left [ (m^2)_i^j (m^2)_j^k  + (m^2)_i^j \mu_{jr} \mu^{kr} + \mu_{ir} \mu^{jr} (m^2)_j^k + \mu_{ij} (m^2)_r^j \mu^{kr} + b_{ij} b^{jk} \right ] 
\nonumber \\ &&
+ y^{ikm} y_{jln} y^{npq} y_{mpq} (m^2)_k^l \left [ \frac{1}{2} (m^2)_i^j + 2 \mu_{ir} \mu^{jr} \right ]  
+ y^{imn} y_{jmp} y^{kqp} y_{lqn} (m^2)_k^l \left [ 3 (m^2)_i^j  + 4 \mu_{ir} \mu^{jr} \right ] 
\nonumber \\ &&
-\frac{1}{2} y^{ilm} y_{knp} y_{lmq} y^{npq} \left [ (m^2)_i^j (m^2)_j^k + (m^2)_i^j \mu_{jr} \mu^{kr} + \mu_{ir} \mu^{jr} (m^2)_j^k +
\mu_{ij} (m^2)_r^j \mu^{kr}  + b_{ij} b^{jk} \right ]
\nonumber \\ &&
-\frac{1}{4} y^{imn} y_{lmn} y_{jpq} y^{kpq} (m^2)_k^l \left [ (m^2)_i^j + 2 \mu_{ir} \mu^{jr} \right ] 
- y^{ikl} y_{jkm} y^{nqr} y_{pqr} \mu_{ln} \mu^{mp} (m^2)_i^j 
\nonumber \\ &&
-  y^{ikm} y_{jkl} (m^2)_i^j \left [ \mu_{mr} \mu^{nr} y_{npq} y^{lpq} + \mu_{nr} \mu^{lr} y_{mpq} y^{npq} \right ]
\nonumber \\ &&
+ 2  y^{ikm} y_{jkl} \mu_{ir} \mu^{jr}  \left [ (m^2)_m^n y_{npq} y^{lpq} + (m^2)_n^l y_{mpq} y^{npq} \right ]
\nonumber \\ &&
-\frac{1}{2} y^{ijk} y_{ijl} y^{mnp} y_{mnq}
\left [\mu^{lq} \mu_{kr} (m^2)^r_p + \mu_{kp} \mu^{lr} (m^2)_r^q  
\right ]
\nonumber \\ &&
+ 12 \zeta_3 y^{ijk} y^{lpn} y_{lmr} y_{knq} (m^2)_i^r \mu_{jp} \mu^{mq} 
+ 6 \zeta_3 y^{ijk} y^{lmn} y_{imp} y_{knq} \left [ \mu_{jl} \mu^{qr} (m^2)_r^p + \mu^{qp} \mu_{jr} (m^2)_l^r \right ]
\nonumber \\ &&
+ 2  \left [ a^{ikl} a_{jkm} y^{mpq} y_{lpq} + y^{ikl} y_{jkm} a^{mpq} a_{lpq}
           + a^{ikl} y_{jkm} y^{mpq} a_{lpq} + y^{ikl} a_{jkm} a^{mpq} y_{lpq} \right ] \mu_{ir} \mu^{jr}
\nonumber \\ &&
- \frac{1}{2} \left [ a^{ijk} a_{ijl} y^{lmn} y_{pmn} + a^{ijk} y_{ijl} y^{lmn} a_{pmn} 
                      + y^{ijk} a_{ijl} a^{lmn} y_{pmn} + y^{ijk} y_{ijl} a^{lmn} a_{pmn} \right] \mu_{kr} \mu^{pr} 
\nonumber \\ &&
-\frac{1}{2} \left [ a^{ijk} a_{ijl} y^{mpq} y_{npq} + 
                     a^{ijk} y_{ijl} y^{mpq} a_{npq}\right ] \mu^{ln} \mu_{km} 
-\frac{1}{4} y^{imn} y_{kmn} y^{jpq} y_{lpq} b_{ij} b^{kl}
\nonumber \\ &&
+ 2 \left [ a^{ikl} a_{jkm} y^{mpq} y_{lpq}
          + a^{ikl} y_{jkm} y^{mpq} a_{lpq}
          + y^{ikl} a_{jkm} a^{mpq} y_{lpq} \right ] (m^2)_i^j 
\nonumber \\ &&
- \frac{1}{2} \left [ a^{ijk} a_{ijl} y^{lmn} y_{pmn} + a^{ijk} y_{ijl} y^{lmn} a_{pmn} 
                    + y^{ijk} y_{ijl} a^{lmn} a_{pmn} \right] (m^2)_k^p 
\nonumber \\ &&
+ \left [2 y^{klm} a_{iln} y^{npq} y_{mpq} + 2 y^{klm} y_{iln} y^{npq} a_{mpq}
  -\frac{1}{2} y_{imn} y^{kpq} y^{mnl} a_{pql} -\frac{1}{2} a_{imn} y^{kpq} y^{mnl} y_{pql}
                    \right ] b^{ij} \mu_{jk} 
\nonumber \\ &&
+ \left [2 y_{klm} a^{iln} y_{npq} y^{mpq} + 2 y_{klm} y^{iln} y_{npq} a^{mpq} 
  -\frac{1}{2} y^{imn} y_{kpq} y_{mnl} a^{pql} -\frac{1}{2} a^{imn} y_{kpq} y_{mnl} y^{pql}
                    \right ] b_{ij} \mu^{jk} 
\nonumber \\ &&
-\frac{1}{2} \left [y^{ijk} a_{ijl} b^{lm} \mu_{kn} y^{npq} y_{mpq}  
+ y_{ijk} a^{ijl} b_{lm} \mu^{kn} y_{npq} y^{mpq} \right ]
-\frac{1}{4} a^{ijk} a_{ijl} a^{lmn} a_{kmn}
\nonumber \\ &&
+ 3 \zeta_3 \left [\bigl (y^{ijk} b^{lm} + 2 a^{ijk} \mu^{lm}\bigr )
\bigl (y_{iln} b_{jp} + 2 a_{iln} \mu_{jp} \bigr ) y_{kmq} y^{npq} \right ]
,
\label{eq:betaL3a}
\eeq
\vspace{-1cm}
\beq
\beta_\Lambda^{(3b)} &=& 
6 g_a^2 C_a(i) a^{ijk} a_{ijk} |M_a|^2 +
g_a^2 a^{ijk} a_{ijl} (m^2)_k^l \left [(12 \zeta_3 - 8) C_a(i) + (4 - 18 \zeta_3) C_a(k) \right ]  
\nonumber \\ &&
+ g_a^2 \left (y^{ijk} a_{ijl} M_a + a^{ijk} y_{ijl} M_a^* \right ) (m^2)_k^l \left [(8 - 12 \zeta_3) C_a(i) +
(6 \zeta_3 - 6) C_a(k) \right ]
\nonumber \\ &&
+ 5 g_a^2 C_a(i) \left (y^{ijk} a_{ijk} M_a + a^{ijk} y_{ijk} M_a^* \right ) |M_a|^2
- 10 g_a^2 y^{ijk} y_{ijk} C_a(i) |M_a|^4
\nonumber \\ &&
+ g_a^2 y^{ijk} y_{ijl} (m^2)_k^l |M_a|^2 \left [14 C_a(k) -4 C_a(i) \right ]
\nonumber \\ &&
+ g_a^2 y^{ijk} y_{ijl} (m^2)_k^n (m^2)_n^l \left [(12 \zeta_3 - 8) C_a(i) + (8 - 18 \zeta_3) C_a(k) \right ]
\nonumber \\ &&
+ g_a^2 y^{ijk} y_{iln} (m^2)_j^l (m^2)_k^n \left [(12 \zeta_3 - 2) C_a(i) + (12 - 24 \zeta_3) C_a(j) \right ]
\nonumber \\ &&
+ g_a^2 \left [ (12 \zeta_3 - 8) C_a(i) + (10 - 18 \zeta_3) C_a(k) \right ] \Bigl \{
\bigl (a^{ijk} - y^{ijk} M_a \bigr ) \bigl ( a_{ijl} - y_{ijl} M_a^* \bigr ) \mu_{kn} \mu^{ln}
\nonumber \\ &&
+ y^{ijk} y_{ijl} \Bigl [ 
\bigl ( b^{nl} - \mu^{nl} M_a \bigr )\bigl (b_{kn} -  \mu_{kn} M_a^*\bigr )
+ (m^2)_k^n \mu_{np} \mu^{lp} + \mu_{kp} \mu^{np} (m^2)_n^l  + \mu_{kn} (m^2)_p^n \mu^{pl} \Bigr ] 
\nonumber \\ &&
+ \left [ y^{ijk} y_{iln} (m^2)_j^l  
+ y^{ijk} y_{jln} (m^2)_i^l \right ] \mu_{kp} \mu^{np} 
+ y^{ijk} a_{ijl} \mu_{kn} b^{ln} 
+ a^{ijk} y_{ijl} b_{kn} \mu^{ln}
\Bigr \}
,
\label{eq:betaL3b}
\\
\beta_\Lambda^{(3c)} &=& 
g_a^4 d_a  \left [ (4 - 24 \zeta_3) G_a^2 + (96 \zeta_3 -26) G_a S_a + 10 S_a^2\right ] |M_a|^4
\nonumber \\ &&
+ g_a^2 g_b^2 d_a S_{ab} |M_a|^2 \left [
(20 - 24 \zeta_3)  (2 |M_a|^2 + M_a M_b^* + M_a^* M_b) - 24 |M_b|^2 
\right ]
\nonumber \\ &&
+ g_a^2 C_a(i) \left \{ 
g_a^2 \left [ (84 - 120 \zeta_3) G_a - 36 S_a \right ]
+ g_b^2 C_b(i)  (96 \zeta_3 - 64) 
\right \} (m^2)_i^i  |M_a|^2
\nonumber \\ &&
+ g_a^2 g_b^2 C_a(i) C_b(i) \left [(48 \zeta_3 - 40)  (m^2)_i^i  + (48\zeta_3 - 56) \mu^{ij} \mu_{ij} \right] M_a M_b^* 
\nonumber \\ &&
+ g_a^2 C_a(i)  \left [ g_a^2 (18 G_a - 6 S_a) - 8 g_b^2 C_b(i) \right ] (m^2)_i^j (m^2)_j^i
\nonumber \\ &&
+ g_a^4 C_a(i) \left[ (56 - 24 \zeta_3) G_a - 16 S_a \right ] \mu^{ij} \mu_{ij} |M_a|^2  
\nonumber \\ &&
+ g_a^2 C_a(i) 
\Bigl [(24 - 12 \zeta_3) g_a^2 G_a -8 g_a^2 S_a + (24 \zeta_3 - 28) g_b^2 C_b(i) \Bigr ]
\!
\Bigl [2 (m^2)_i^j \mu_{jk} \mu^{ik} 
\nonumber \\ && 
+ (b^{ij} - 2 M_a \mu^{ij})(b_{ij} - 2 M_a^* \mu_{ij}) \Bigr ] 
+ 8 g_a^4 (m^2)_i^i \left [ (m^2)_j^j - \mu^{jk} \mu_{jk} \right ] C_a(i) C_a(j)/d_a
.
\label{eq:betaL3c}
\eeq
Note that $\beta_\Lambda$ vanishes in the case of no supersymmetry breaking terms.

To obtain the 3-loop beta function for $\Lambda$, I found that it was more than sufficient to consider 
eq.~(\ref{eq:RGcheck}) for the following example models, chosen somewhat arbitrarily.
Since the goal here was only to obtain the beta function for the field-independent vacuum energy, the background values of all scalar fields were simply set to 0.
\begin{itemize}
\item Supersymmetric theory with no gauge symmetry and 6 chiral superfields $\Phi_{1,2,3,4,5,6}$, with superpotential
$ W = y \Phi_1 \Phi_2 \Phi_3 + y' \Phi_1 \Phi_4 \Phi_5 + y'' \Phi_2 \Phi_4 \Phi_6 
+ \tfrac{1}{2} \mu_1 \Phi_1^2
+ \tfrac{1}{2} \mu_2 \Phi_2^2
+ \tfrac{1}{2} \mu_3 \Phi_3^2
+ \tfrac{1}{2} \mu_4 \Phi_4^2
+ \tfrac{1}{2} \mu_5 \Phi_5^2
+ \tfrac{1}{2} \mu_5 \Phi_6^2
$
and soft supersymmetry-breaking Lagrangian
$-{\cal L}_{\rm soft} = 
m^2_1 |\phi_1|^2 + m^2_2 |\phi_2|^2 + m^2_3 |\phi_3|^2 + m^2_4 |\phi_4|^2 
+ m^2_5 |\phi_5|^2 + m^2_6 |\phi_6|^2 + \Lambda$.
\item Supersymmetric theory with no gauge symmetry and 6 chiral superfields $\Phi_{1,2,3,4,5,6}$, with superpotential
$W = y \Phi_1 \Phi_2 \Phi_3 + y' \Phi_1 \Phi_4 \Phi_5 + y'' \Phi_2 \Phi_4 \Phi_6 + 
\mu_{12} \Phi_1 \Phi_2 + \mu_{34} \Phi_3 \Phi_4 + \mu_{56} \Phi_5 \Phi_6$ and soft supersymmetry-breaking Lagrangian
$-{\cal L}_{\rm soft} = m^2_1 |\phi_1|^2 + m^2_2 |\phi_2|^2 + m^2_3 |\phi_3|^2 + m^2_4 |\phi_4|^2 
+ m^2_5 |\phi_5|^2 + m^2_6 |\phi_6|^2 + \Lambda$.
\item Supersymmetric theory with no gauge symmetry and 2 chiral superfields, with the most general superpotential (including four independent Yukawa couplings and three mass terms) and the most general soft supersymmetry-breaking Lagrangian.
\item Supersymmetric $U(1)$ gauge theory with 6 chiral superfields with charges $q_1, -q_1, q_2, -q_2, 0$, and $0$, with the most general allowed superpotential and the most general soft supersymmetry-breaking Lagrangian consistent with these charge assignments.
\item Supersymmetric $U(1)$ gauge theory with 6 chiral superfields with charges $q_1, -q_1, q_2, -q_2, (q_1+q_2)$, and $-(q_1+q_2)$, with the most general allowed superpotential and soft supersymmetry-breaking Lagrangian consistent with these charge assignments.
\item Supersymmetric $SU(2)$ gauge theory, with three doublet chiral superfields $\Phi_1$, $\Phi_2$, and $\Phi_3$
and one singlet $S$, with superpotential $W = y \Phi_1 \Phi_2 S$, and soft supersymmetry breaking Lagrangian
$-{\cal L}_{\rm soft} = \left (a \phi_1 \phi_2 s  + \tfrac{1}{2} M \lambda\lambda \right ) + {\rm c.c.} 
+ m^2_1 |\phi_1|^2 + m^2_2 |\phi_2|^2 + m^2_3 |\phi_3|^2 + m^2_s |s|^2 + \Lambda$.
\item Supersymmetric $SU(2)$ gauge theory, with four doublet chiral superfields $\Phi_1$, $\Phi_2$, $\Phi_3$, $\Phi_4$
and one singlet $S$, with superpotential $W = y \Phi_1 \Phi_2 S + y' \Phi_3 \Phi_4 S$, and soft supersymmetry breaking Lagrangian
$-{\cal L}_{\rm soft} = \left (a \phi_1 \phi_2 s  + a' \phi_3 \phi_4 s + \tfrac{1}{2} M \lambda\lambda \right ) + {\rm c.c.} 
+ m^2_1 |\phi_1|^2 + m^2_2 |\phi_2|^2 + m^2_3 |\phi_3|^2 + m^2_4 |\phi_4|^2  + m^2_s |s|^2 + \Lambda$.
\item Supersymmetric $SU(2) \times U(1)$ gauge theory, with chiral superfields
transforming as $({\bf 2}, +1)$, $({\bf 2}, -1)$, and $({\bf 1}, 0)$, and the most general allowed superpotential
and soft supersymmetry-breaking Lagrangian.
\item Supersymmetric $SU(3)$ gauge theory, with triplet and anti-triplet chiral superfields $\Phi$, $\overline \Phi$ and
and one singlet $S$, with superpotential $W = y \Phi \overline \Phi S$, and soft supersymmetry breaking Lagrangian
$-{\cal L}_{\rm soft} = \left (a \phi \overline \phi s  + \tfrac{1}{2} M \lambda\lambda \right ) + {\rm c.c.} 
+ m^2 |\phi|^2 + \overline{m}^2 |\overline \phi|^2 + m^2_s |s|^2 + \Lambda$.
\end{itemize}
The expression for $\beta^{(3)}_\Lambda$ in eqs.~(\ref{eq:betaL3})-(\ref{eq:betaL3c}) was obtained by writing the most general possible form for it with unknown coefficients, and then solving for the coefficients by demanding the vanishing of eq.~(\ref{eq:RGcheck}) for $\ell = 1,2,3$.  These examples also produced numerous redundant checks.

To obtain the previously unknown values of the 3-loop scalar field anomalous dimension coefficients $n_7,\ldots n_{26}$ in eq.~(\ref{eq:coeffsgammaS}), I found that
it was again more than sufficient to consider eq.~(\ref{eq:RGcheck}) for each of the following example models:
\begin{itemize}
\item Supersymmetric $U(1)$ gauge theory with six chiral superfields $\Phi_{1,2,3,4,5,6}$, with charges $q_1,\, q_2,\, q_3,\, q_4,\, -(q_1+q_2),$ and $-(q_1 + q_3)$, and a superpotential $W = y \Phi_1 \Phi_2 \Phi_5 + y' \Phi_1 \Phi_3 \Phi_6$. To avoid a gauge anomaly, $q_4^3 = q_1^3 + 3 q_1^2 q_2 + 3 q_1 q_2^2 + 3 q_1^2 q_3 + 3 q_1 q_3^2$.
The effective potential is a function of the Yukawa couplings $y$, $y'$, the gauge coupling $g$, and the background values of the scalar components of $\Phi_1$ and $\Phi_2$, which are taken to be independent.
\item Supersymmetric $SU(2) \times U(1)$ gauge theory with chiral superfields transforming as
$\Phi_1 = ({\bf 2}, q_1)$, 
$\Phi_2 = ({\bf 2}, q_2)$,
$\Phi_3 = ({\bf 2}, q_3)$,
$\Phi_4 = ({\bf 1}, q_4)$,
$\Phi_5 = ({\bf 1}, -q_1-q_2)$.
There is a Yukawa interaction $W = y \Phi_1 \Phi_2 \Phi_5$.
To avoid gauge anomalies, $q_3 = -q_1 - q_2$, and $q_4^3 = q_1^3 + 9 q_1^2 q_2 + 9 q_1 q_2^2 + q_2^3$.
The effective potential is a function of the Yukawa coupling $y$, the gauge couplings $g$ and $g'$, and the background values of the scalar components of $\Phi_1$ and $\Phi_2$, which are taken to be independent.
\item Supersymmetric $SU(2)$ gauge theory with chiral superfields consisting of one doublet $\Phi$ and one triplet
$\Sigma$, with superpotential $W = y \Phi\Phi\Sigma$. 
The effective potential is a function of the Yukawa coupling $y$, the gauge couplings $g$, and the background values of the scalar components $\Phi$ and $\Sigma$, which are taken to be independent.
\item Supersymmetric $U(1)$, $SU(2)$, and $SU(3)$ gauge theories with chiral superfields $\Phi$ and $\overline \Phi$ in fundamental and anti-fundamental representations, with superpotential $W = \mu \Phi \overline \Phi$ and $-{\cal L}_{\rm soft} = \Lambda + m^2 |\phi|^2 + \overline m^2 |\overline \phi|^2 +
(b \phi \overline \phi + \tfrac{1}{2} M \lambda^a \lambda^a + {\rm c.c.})$. The background scalar field components of $\phi$ and $\overline\phi$ were taken to be non-zero and equal along a $D$-flat direction,
but supersymmetry is explicitly broken by soft terms, so the effective potential does not vanish.
\end{itemize}
In addition to determining the scalar field anomalous dimension coefficients, these models again produced numerous redundant checks
of eq.~(\ref{eq:RGcheck}).

\section{Outlook\label{sec:outlook}}
\setcounter{equation}{0}
\setcounter{figure}{0}
\setcounter{table}{0} 
\setcounter{footnote}{1}

In this paper, I have provided the 3-loop effective potential in Landau gauge for a general softly broken supersymmetric theory, using a regularization and renormalization scheme that respects supersymmetry. As byproducts, the beta function for the field-independent vacuum energy and the Landau gauge anomalous dimension of scalars were obtained. 
 
It should be noted that the results obtained in this paper apply {\em only} to models with softly broken supersymmetry. This is because if there is supersymmetry violation in the dimensionless couplings (or simply in the field content) of the theory, then it was shown in ref.~\cite{Jack:1993ws,Jack:1994bn} that while dimensional reduction can be applied in a consistent way,  renormalization requires that there are evanescent couplings that are different (at all but at most one renormalization scale) for $\epsilon$-scalars and vectors. This is inconsistent with the procedure followed in the present paper, where the contributions of $\epsilon$-scalars and vectors have been combined due to always having the same gauge interactions. This is a feature only of softly broken supersymmetry.

I have checked the consistency of the three-loop effective potential for numerous toy models, as described above. An obvious more practical application of the results obtained here is to the MSSM, which could well describe our world even though there are increasingly stringent bounds on superpartners coming from direct searches at the Large Hadron Collider. This would extend the 2-loop results of ref.~\cite{Martin:2002iu}, and allow a more  precise determination of the relations between the Higgs vacuum expectation values and the other renormalized Lagrangian parameters. Implementing the general results found here in the special case of the MSSM is in principle straightforward, although the combinatorics appear to be somewhat intimidating. This is left as an exercise for the clever and courageous reader. 

\vspace{0.3cm}

This work is supported in part by the National Science Foundation grants with award numbers 2013340 and 2310533.



\begin{thebibliography}{90}
\baselineskip=14.4pt

\bibitem{Coleman:1973jx}
  S.~R.~Coleman and E.~J.~Weinberg,
  ``Radiative Corrections as the Origin of Spontaneous Symmetry Breaking,''
  Phys.\ Rev.\ D {\bf 7}, 1888 (1973).

\bibitem{Jackiw:1974cv}
  R.~Jackiw,
  ``Functional evaluation of the effective potential,''
  Phys.\ Rev.\ D {\bf 9}, 1686 (1974).

\bibitem{Sher:1988mj}
  M.~Sher,
  ``Electroweak Higgs Potentials and Vacuum Stability,''
  Phys.\ Rept.\  {\bf 179}, 273 (1989),
  and references therein.

\bibitem{Ford:1992pn}
  C.~Ford, I.~Jack and D.R.T.~Jones,
  ``The Standard model effective potential at two loops,''
  Nucl.\ Phys.\ B {\bf 387}, 373 (1992)
  [Erratum-ibid.\ B {\bf 504}, 551 (1997)]
  [hep-ph/0111190],

\bibitem{Martin:2001vx}
  S.P.~Martin,
  ``Two loop effective potential for a general renormalizable theory
  and softly broken supersymmetry,''
  Phys.\ Rev.\ D {\bf 65}, 116003 (2002)
  [hep-ph/0111209].

\bibitem{Martin:2013gka}
  S.~P.~Martin,
  ``Three-loop Standard Model effective potential at leading order in 
  strong and top Yukawa couplings,''
  Phys.\ Rev.\ D {\bf 89}, no. 1, 013003 (2014)
  [arXiv:1310.7553 [hep-ph]].

\bibitem{Martin:2017lqn}
S.~P.~Martin,
``Effective potential at three loops,''
Phys. Rev. D \textbf{96}, no.9, 096005 (2017)
[arXiv:1709.02397 [hep-ph]].

\bibitem{Martin:2018emo}
S.~P.~Martin and H.~H.~Patel,
``Two-loop effective potential for generalized gauge fixing,''
Phys. Rev. D \textbf{98}, no.7, 076008 (2018)
[arXiv:1808.07615 [hep-ph]].

\bibitem{Martin:2015eia}
  S.~P.~Martin,
  ``Four-loop Standard Model effective potential at leading order in QCD,''
  Phys.\ Rev.\ D {\bf 92}, no. 5, 054029 (2015)
  [arXiv:1508.00912 [hep-ph]].

\bibitem{Bollini:1972ui}
C.~G.~Bollini and J.~J.~Giambiagi,
``Dimensional Renormalization: The Number of Dimensions as a Regularizing Parameter,''
Nuovo Cim. B \textbf{12}, 20-26 (1972)
doi:10.1007/BF02895558

\bibitem{Bollini:1972bi}
C.~G.~Bollini and J.~J.~Giambiagi,
``Lowest order divergent graphs in nu-dimensional space,''
Phys. Lett. B \textbf{40}, 566-568 (1972)
doi:10.1016/0370-2693(72)90483-2

\bibitem{Ashmore:1972uj}
J.~F.~Ashmore,
``A Method of Gauge Invariant Regularization,''
Lett. Nuovo Cim. \textbf{4}, 289-290 (1972)
doi:10.1007/BF02824407

\bibitem{Cicuta:1972jf}
G.~M.~Cicuta and E.~Montaldi,
``Analytic renormalization via continuous space dimension,''
Lett. Nuovo Cim. \textbf{4}, 329-332 (1972)
doi:10.1007/BF02756527

\bibitem{tHooft:1972tcz}
G.~'t Hooft and M.~J.~G.~Veltman,
``Regularization and Renormalization of Gauge Fields,''
Nucl. Phys. B \textbf{44}, 189-213 (1972)
doi:10.1016/0550-3213(72)90279-9

\bibitem{tHooft:1973mfk}
G.~'t Hooft,
``Dimensional regularization and the renormalization group,''
Nucl. Phys. B \textbf{61}, 455-468 (1973)
doi:10.1016/0550-3213(73)90376-3

\bibitem{Bardeen:1978yd}
W.~A.~Bardeen, A.~J.~Buras, D.~W.~Duke and T.~Muta,
``Deep Inelastic Scattering Beyond the Leading Order in Asymptotically Free Gauge Theories,''
Phys. Rev. D \textbf{18}, 3998 (1978)
doi:10.1103/PhysRevD.18.3998

\bibitem{Braaten:1981dv}
E.~Braaten and J.~P.~Leveille,
``Minimal Subtraction and Momentum Subtraction in {QCD} at Two Loop Order,''
Phys. Rev. D \textbf{24}, 1369 (1981)
doi:10.1103/PhysRevD.24.1369

\bibitem{Siegel:1979wq}
W.~Siegel,
``Supersymmetric Dimensional Regularization via Dimensional Reduction,''
Phys. Lett. B \textbf{84}, 193-196 (1979)
doi:10.1016/0370-2693(79)90282-X

\bibitem{Capper:1979ns}
D.~M.~Capper, D.~R.~T.~Jones and P.~van Nieuwenhuizen,
``Regularization by Dimensional Reduction of Supersymmetric and Nonsupersymmetric Gauge Theories,''
Nucl. Phys. B \textbf{167}, 479-499 (1980)
doi:10.1016/0550-3213(80)90244-8

\bibitem{Jack:1997sr}
I.~Jack and D.~R.~T.~Jones,
``Regularization of supersymmetric theories,''
Adv. Ser. Direct. High Energy Phys. \textbf{21}, 494-513 (2010)
[arXiv:hep-ph/9707278 [hep-ph]].

\bibitem{Siegel:1980qs}
W.~Siegel,
``Inconsistency of Supersymmetric Dimensional Regularization,''
Phys. Lett. B \textbf{94}, 37-40 (1980)
doi:10.1016/0370-2693(80)90819-9

\bibitem{Avdeev:1982xy}
L.~V.~Avdeev and A.~A.~Vladimirov,
``Dimensional Regularization and Supersymmetry,''
Nucl. Phys. B \textbf{219}, 262-276 (1983)
doi:10.1016/0550-3213(83)90437-6

\bibitem{Stockinger:2005gx}
D.~Stockinger,
``Regularization by dimensional reduction: consistency, quantum action principle, and supersymmetry,''
JHEP \textbf{03}, 076 (2005)
[arXiv:hep-ph/0503129 [hep-ph]].

\bibitem{Jack:1994rk}
I.~Jack, D.~R.~T.~Jones, S.~P.~Martin, M.~T.~Vaughn and Y.~Yamada,
``Decoupling of the epsilon scalar mass in softly broken supersymmetry,''
Phys. Rev. D \textbf{50}, R5481-R5483 (1994)
[arXiv:hep-ph/9407291 [hep-ph]].

\bibitem{Jack:1994kd}
I.~Jack and D.~R.~T.~Jones,
``Soft supersymmetry breaking and finiteness,''
Phys. Lett. B \textbf{333}, 372-379 (1994)
[arXiv:hep-ph/9405233 [hep-ph]].

\bibitem{Martin:2016bgz}
S.~P.~Martin and D.~G.~Robertson,
``Evaluation of the general 3-loop vacuum Feynman integral,''
Phys. Rev. D \textbf{95}, no.1, 016008 (2017)
[arXiv:1610.07720 [hep-ph]].
The {\tt 3VIL} code is available at\\
\href{https://davidgrobertson.github.io/3VIL/}{https://davidgrobertson.github.io/3VIL/}

\bibitem{Broadhurst:1991fi}
  D.~J.~Broadhurst,
  ``Three loop on-shell charge renormalization without integration: Lambda-MS (QED) to four loops,''
  Z.\ Phys.\ C {\bf 54}, 599 (1992).

\bibitem{Avdeev:1994db}
  L.~Avdeev, J.~Fleischer, S.~Mikhailov and O.~Tarasov,
  ``$O (\alpha \alpha_s^2)$ correction to the electroweak rho parameter,''
  Phys.\ Lett.\ B {\bf 336}, 560 (1994)
  [Phys.\ Lett.\ B {\bf 349}, 597 (1995)]
  [hep-ph/9406363].

\bibitem{Fleischer:1994dc}
  J.~Fleischer and O.~V.~Tarasov,
  ``Application of conformal mapping and Pad\'e approximants $(\omega P' s)$ to the calculation of various two-loop Feynman diagrams,''
  Nucl.\ Phys.\ Proc.\ Suppl.\  {\bf 37B}, no. 2, 115 (1994)
  [hep-ph/9407235].

\bibitem{Avdeev:1995eu}
  L.~V.~Avdeev,
  ``Recurrence relations for three loop prototypes of bubble diagrams with a mass,''
  Comput.\ Phys.\ Commun.\  {\bf 98}, 15 (1996)
  [hep-ph/9512442].

\bibitem{Broadhurst:1998rz}
  D.~J.~Broadhurst,
  ``Massive three-loop Feynman diagrams reducible to SC* primitives of
  algebras of the sixth root of unity,''
  Eur.\ Phys.\ J.\ C {\bf 8}, 311 (1999)
  [hep-th/9803091].

\bibitem{Fleischer:1999mp}
  J.~Fleischer and M.~Y.~Kalmykov,
  ``Single mass scale diagrams: Construction of a basis for the epsilon expansion,''
  Phys.\ Lett.\ B {\bf 470}, 168 (1999)
  [hep-ph/9910223].

\bibitem{Schroder:2005va}
  Y.~Schr\"oder and A.~Vuorinen,
  ``High-precision epsilon expansions of single-mass-scale
  four-loop vacuum bubbles,''
  JHEP {\bf 0506}, 051 (2005)
  [hep-ph/0503209].

\bibitem{Martin:2014bca}
S.~P.~Martin,
``Taming the Goldstone contributions to the effective potential,''
Phys. Rev. D \textbf{90}, no.1, 016013 (2014)
[arXiv:1406.2355 [hep-ph]].

\bibitem{Elias-Miro:2014pca}
J.~Elias-Miro, J.~R.~Espinosa and T.~Konstandin,
``Taming Infrared Divergences in the Effective Potential,''
JHEP \textbf{08}, 034 (2014)
[arXiv:1406.2652 [hep-ph]].

\bibitem{Pilaftsis:2015bbs}
  A.~Pilaftsis and D.~Teresi,
  ``Symmetry-Improved 2PI Approach to the Goldstone-Boson IR Problem
    of the SM Effective Potential,''
  Nucl.\ Phys.\ B {\bf 906}, 381 (2016)
  [1511.05347 [hep-ph]].

\bibitem{Kumar:2016ltb}
N.~Kumar and S.~P.~Martin,
``Resummation of Goldstone boson contributions to the MSSM effective potential,''
Phys. Rev. D \textbf{94}, no.1, 014013 (2016)
[arXiv:1605.02059 [hep-ph]].

\bibitem{Espinosa:2016uaw}
J.~R.~Espinosa, M.~Garny and T.~Konstandin,
``Interplay of Infrared Divergences and Gauge-Dependence of the Effective Potential,''
Phys. Rev. D \textbf{94}, no.5, 055026 (2016)
[arXiv:1607.08432 [hep-ph]].

\bibitem{Braathen:2016cqe}
J.~Braathen and M.~D.~Goodsell,
``Avoiding the Goldstone Boson Catastrophe in general renormalisable field theories at two loops,''
JHEP \textbf{12}, 056 (2016)
[arXiv:1609.06977 [hep-ph]].

\bibitem{Braathen:2017izn}
J.~Braathen, M.~D.~Goodsell and F.~Staub,
``Supersymmetric and non-supersymmetric models without catastrophic Goldstone bosons,''
Eur. Phys. J. C \textbf{77}, no.11, 757 (2017)
[arXiv:1706.05372 [hep-ph]].

\bibitem{Martin:1997ns}
S.~P.~Martin,
``A Supersymmetry primer,''
[arXiv:hep-ph/9709356 [hep-ph]].

\bibitem{Dreiner:2023yus}
H.~K.~Dreiner, H.~E.~Haber and S.~P.~Martin,
``From Spinors to Supersymmetry,''
Cambridge University Press, 2023,
doi:10.1017/9781139049740

\bibitem{Zumino:1974bg}
B.~Zumino,
``Supersymmetry and the Vacuum,''
Nucl. Phys. B \textbf{89}, 535 (1975)
doi:10.1016/0550-3213(75)90194-7

\bibitem{West:1984dg}
P.~C.~West,
``The Yukawa beta Function in N=1 Rigid Supersymmetric Theories,''
Phys. Lett. B \textbf{137}, 371-373 (1984)
doi:10.1016/0370-2693(84)91734-9

\bibitem{Jack:1996qq}
I.~Jack, D.~R.~T.~Jones and C.~G.~North,
``N=1 supersymmetry and the three loop anomalous dimension for the chiral superfield,''
Nucl. Phys. B \textbf{473}, 308-322 (1996)
[arXiv:hep-ph/9603386 [hep-ph]].

\bibitem{Jones:1974pg}
D.~R.~T.~Jones,
``Asymptotic Behavior of Supersymmetric Yang-Mills Theories in the Two Loop Approximation,''
Nucl. Phys. B \textbf{87}, 127 (1975)
doi:10.1016/0550-3213(75)90256-4

\bibitem{Jones:1983vk}
D.~R.~T.~Jones and L.~Mezincescu,
``The Beta Function in Supersymmetric {Yang-Mills} Theory,''
Phys. Lett. B \textbf{136}, 242-244 (1984)
doi:10.1016/0370-2693(84)91154-7

\bibitem{Jack:1996vg}
I.~Jack, D.~R.~T.~Jones and C.~G.~North,
``N=1 supersymmetry and the three loop gauge Beta function,''
Phys. Lett. B \textbf{386}, 138-140 (1996)
[arXiv:hep-ph/9606323 [hep-ph]].
 
\bibitem{Novikov:1985rd}
V.~A.~Novikov, M.~A.~Shifman, A.~I.~Vainshtein and V.~I.~Zakharov,
``The beta function in supersymmetric gauge theories. Instantons versus traditional approach,''
Phys. Lett. B \textbf{166}, 329-333 (1986)
doi:10.1016/0370-2693(86)90810-5

\bibitem{Shifman:1986zi}
M.~A.~Shifman and A.~I.~Vainshtein,
``Solution of the Anomaly Puzzle in SUSY Gauge Theories and the Wilson Operator Expansion,''
Nucl. Phys. B \textbf{277}, 456 (1986)
doi:10.1016/0550-3213(86)90451-7

\bibitem{Martin:1993yx}
S.~P.~Martin and M.~T.~Vaughn,
``Regularization dependence of running couplings in softly broken supersymmetry,''
Phys. Lett. B \textbf{318}, 331-337 (1993)
[arXiv:hep-ph/9308222 [hep-ph]].

\bibitem{Martin:1993zk}
S.~P.~Martin and M.~T.~Vaughn,
``Two loop renormalization group equations for soft supersymmetry breaking couplings,''
Phys. Rev. D \textbf{50}, 2282 (1994)
[erratum: Phys. Rev. D \textbf{78}, 039903 (2008)]
[arXiv:hep-ph/9311340 [hep-ph]].

\bibitem{Yamada:1994id}
Y.~Yamada,
``Two loop renormalization group equations for soft SUSY breaking scalar interactions: Supergraph method,''
Phys. Rev. D \textbf{50}, 3537-3545 (1994)
[arXiv:hep-ph/9401241 [hep-ph]].

\bibitem{Jack:1997pa}
I.~Jack and D.~R.~T.~Jones,
``The Gaugino Beta function,''
Phys. Lett. B \textbf{415}, 383-389 (1997)
[arXiv:hep-ph/9709364 [hep-ph]].

\bibitem{Jack:1997eh}
I.~Jack, D.~R.~T.~Jones and A.~Pickering,
``Renormalization invariance and the soft Beta functions,''
Phys. Lett. B \textbf{426}, 73-77 (1998)
[arXiv:hep-ph/9712542 [hep-ph]].

\bibitem{Jack:2001ew}
I.~Jack, D.~R.~T.~Jones and R.~Wild,
``Gauge singlet renormalization in softly broken supersymmetric theories,''
Phys. Lett. B \textbf{509}, 131-137 (2001)
[arXiv:hep-ph/0103255 [hep-ph]].

\bibitem{Jack:1998iy}
I.~Jack, D.~R.~T.~Jones and A.~Pickering,
``The soft scalar mass beta function,''
Phys. Lett. B \textbf{432}, 114-119 (1998)
[arXiv:hep-ph/9803405 [hep-ph]].

\bibitem{Jack:1998uj}
I.~Jack, D.~R.~T.~Jones and A.~Pickering,
``The Connection between DRED and NSVZ,''
Phys. Lett. B \textbf{435}, 61-66 (1998)
[arXiv:hep-ph/9805482 [hep-ph]].

\bibitem{Jack:2000jr}
I.~Jack and D.~R.~T.~Jones,
``The Fayet-Iliopoulos D term and its renormalization in the MSSM,''
Phys. Rev. D \textbf{63}, 075010 (2001)
[arXiv:hep-ph/0010301 [hep-ph]].

\bibitem{Jack:1993ws}
I.~Jack, D.~R.~T.~Jones and K.~L.~Roberts,
``Dimensional reduction in nonsupersymmetric theories,''
Z. Phys. C \textbf{62}, 161-166 (1994)
[arXiv:hep-ph/9310301 [hep-ph]].

\bibitem{Jack:1994bn}
I.~Jack, D.~R.~T.~Jones and K.~L.~Roberts,
``Equivalence of dimensional reduction and dimensional regularization,''
Z. Phys. C \textbf{63}, 151-160 (1994)
[arXiv:hep-ph/9401349 [hep-ph]].

\bibitem{Martin:2002iu}
S.~P.~Martin,
``Two Loop Effective Potential for the Minimal Supersymmetric Standard Model,''
Phys. Rev. D \textbf{66}, 096001 (2002)
[arXiv:hep-ph/0206136 [hep-ph]].

\end{thebibliography}
\end{document}